\documentclass[aps,prl,10pt,notitlepage,superscriptaddress,floatfix,twocolumn]{revtex4-2}
\usepackage[colorlinks=true,citecolor=blue,urlcolor=blue,linkcolor=blue]{hyperref}
\usepackage[normalem]{ulem}
\usepackage[utf8]{inputenc}
\usepackage[english]{babel}
\usepackage[T1]{fontenc}
\usepackage{mathtools}
\usepackage{mathrsfs}
\usepackage{amsfonts}
\usepackage{bbm}
\usepackage{amsthm}
\usepackage{xcolor}
\usepackage{bbold}
\usepackage{bm}
\usepackage{soul}
\usepackage{layouts}
\usepackage{array}

\begin{document}

\title{How to grow a straight filament}
\author{L. A. Hoffmann}  
\affiliation{School of Engineering and Applied Sciences, Harvard University, Cambridge, Massachusetts 02138, USA}
\author{L. Mahadevan}
\email{lmahadev@g.harvard.edu}
\affiliation{School of Engineering and Applied Sciences, Harvard University, Cambridge, Massachusetts 02138, USA}
\affiliation{Department of Physics, Harvard University, Cambridge, Massachusetts 02138, USA}
\affiliation{Department of Organismic and Evolutionary Biology, Harvard University, Cambridge, Massachusetts 02138, USA}
\date{\today}
\begin{abstract}
How can a growing biological filament remain straight despite stochastic fluctuations in growth? Motivated by filamentary structures that develop reproducibly across biological systems, we study the stability of a noisy, growing elastic filament regulated by feedback. We formulate a minimal model in which growth responds to the filament's strain, curvature, and orientation through local or nonlocal spatiotemporal feedback laws. Linear stability analysis identifies the conditions under which these feedback mechanisms stabilize a straight configuration. In the presence of noise, we show that purely local feedback requires orientation sensing to suppress long-wavelength instabilities, whereas nonlocal feedback allows stabilization through proprioceptive (curvature) sensing alone. Coupling to an elastic substrate further suppresses large-scale fluctuations. Our results establish minimal control strategies that ensure robust straight growth and suggest experimental signatures for identifying the feedback mechanisms underlying morphogenesis.
\end{abstract}

\maketitle

\textit{Introduction}---Despite the well-documented role of fluctuations at cellular scales during embryonic development, tissue morphogenesis is remarkably reproducible on the scale of whole organs or organisms. This suggests that feedback-control mechanisms regulate development to ensure robust morphodynamics~\cite{shraiman2005,hong2018,damavandi2019,naoki2019}. To achieve this, the system must continuously sense its internal state (interoception), its environment (exteroception), or both, and adjust growth accordingly~\cite{legoff2016,chen2021,harmansa2021,valet2022}.
These feedback processes are particularly relevant for slender geometries, such as filamentous and membranous objects, because they are highly susceptible to long-wavelength, low-energy deformations that are easy to excite and thus difficult to control.

\begin{figure}[b]
\centering
\includegraphics[width=\columnwidth]{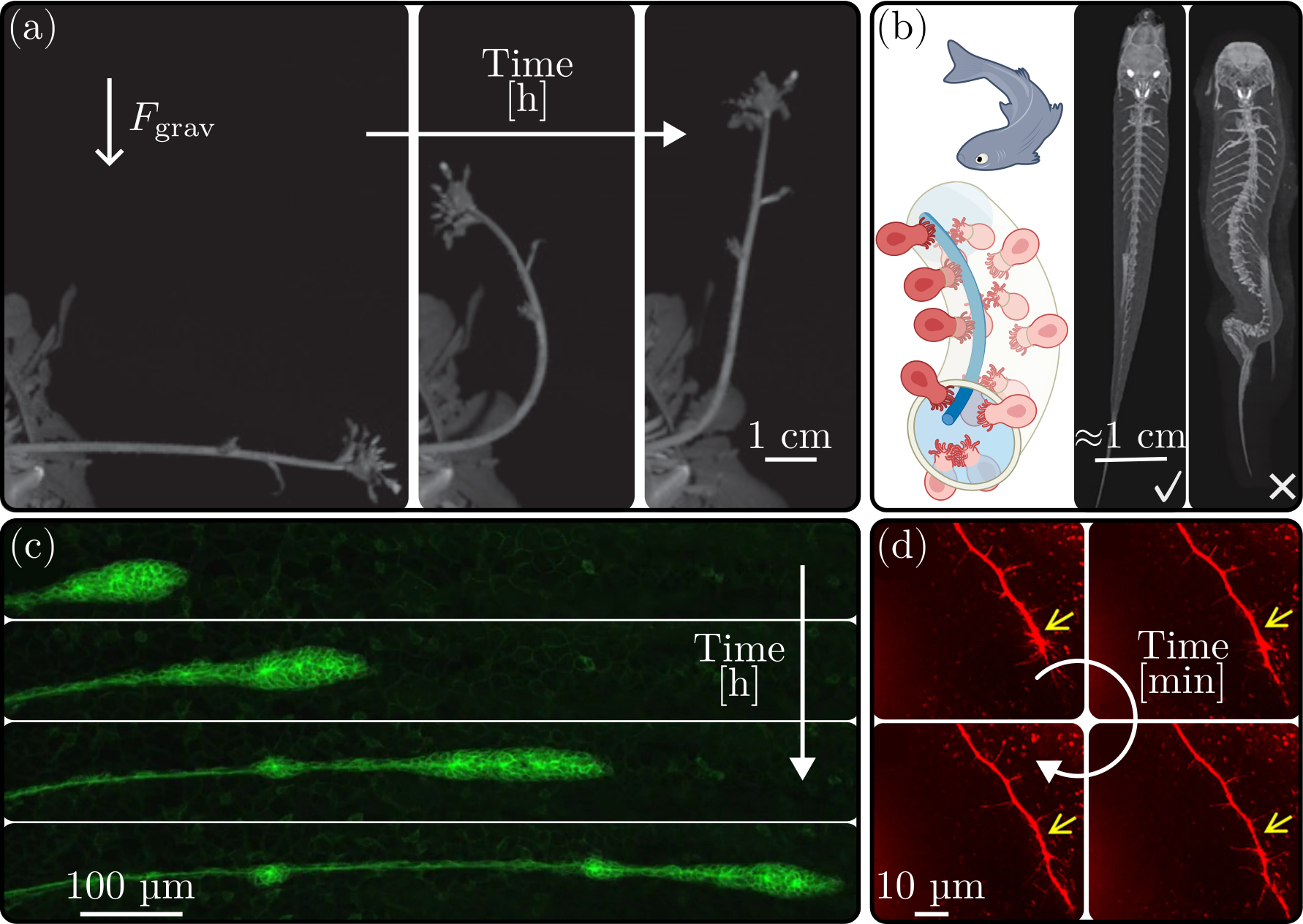} 
\caption{\textbf{Biological examples of straight growth.} (a) Initially horizontal \textit{Arabidopsis} grows to align with the vertical gravitational field. Adapted from Ref.~\cite{bastien2013}. (b) Mechano- and chemosensitive neurons detect spinal curvature during development and are required to guarantee a straight spine, e.g., in zebrafish (right). Adapted from Ref.~\cite{wyart2023} (left) and Ref.~\cite{troutwine2020} (right). (c) Directed collective cell migration forms the lateral line primordium in zebrafish guided by a chemokine signal. Adapted from Ref.~\cite{haas2006}. (d) Axon growth in the \textit{Drosophila} wing is guided by external signals. Adapted from Ref.~\cite{fang2023}.}
\label{fig:Exp}
\end{figure}

Here we focus on the shape and growth of quasi-one-dimensional filamentous structures observed in settings spanning multiple time and length scales, as shown in Fig.~\ref{fig:Exp}. The most common interoceptive mechanisms in this context are \textit{proprioception} (sensing one's own shape) and \textit{strain sensing}. The former has been shown to be required for the straight growth of plant shoots and vertebrate spines~\cite{chelakkot2017,meroz2019,moulia2021,troutwine2020,wyart2023}, as well as for posture control~\cite{blecher2017,bornstein2021,hoffmann2026}; see Fig.~\ref{fig:Exp}(a,b). The latter is known to be relevant in collective cell dynamics~\cite{ladoux2017,hamant2019}.
Complementing these interoceptive mechanisms, external signals can serve as an \textit{orienting} mechanism, establishing a preferred orientation, while mechanical \textit{coupling to a substrate} can define a preferred plane. For example, the gravitational field allows growing plants to distinguish ``up'' and ``down'' (Fig.~\ref{fig:Exp}(a))~\cite{bastien2013,okamoto2015,berut2018}, while a chemical gradient allows directed growth of cells or axons (Fig.~\ref{fig:Exp}(c,d)), whose dynamics are also affected by substrate properties~\cite{haas2006,ladoux2017,stoeckli2018,forghani2023,fang2023}. These sensing mechanisms then dynamically regulate growth, e.g., through growth-hormone signaling~\cite{nakayama2012,hong2018,jonsson2022,xu2024} or cell division~\cite{ladoux2017,moulia2021}, thereby creating a closed feedback loop that allows for tight control of development. A more detailed description in the context of plant growth is given in SM Sec.~SII.

In all cases shown in Fig.~\ref{fig:Exp}, development is impaired in the absence of feedback~\cite{bastien2013,wyart2023,troutwine2020,haas2006,fang2023}, leading to misoriented growth, uncontrolled filament deformations, or both. However, how these mechanisms interact, and what minimal set is required for robust development, remains unclear both experimentally and theoretically. We introduce a model of a noisy, growing one-dimensional elastic filament, implement different feedback laws that capture the aforementioned sensing mechanisms, and analyze their effects on linear stability. This yields a phase diagram linking biological control strategies to stable morphologies.

\begin{figure}
\centering
\includegraphics[width=\columnwidth]{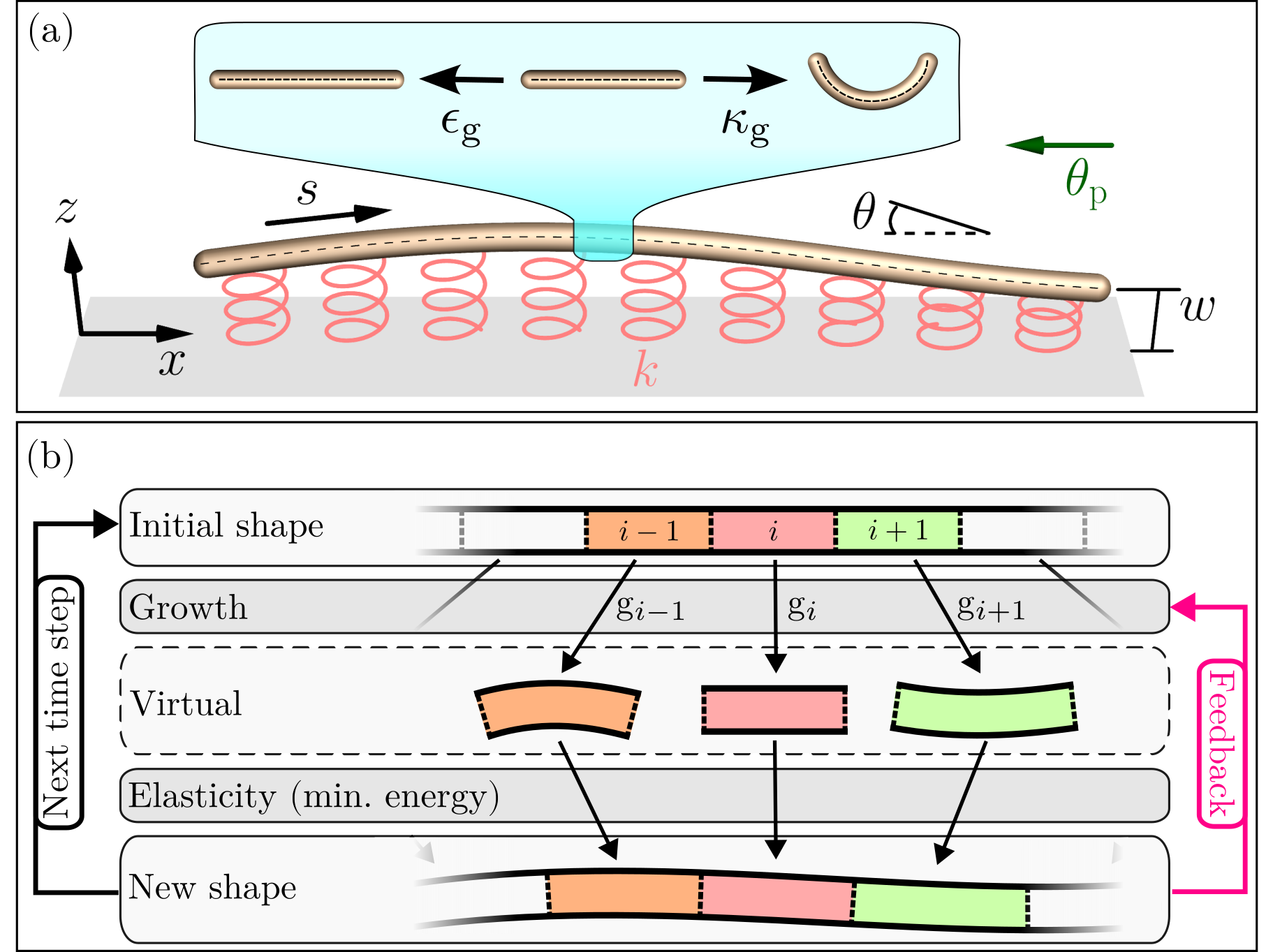} 
\caption{\textbf{Theoretical model.} (a) Sketch of an elastic filament in two-dimensional space $(x,z)$ described by the coordinate $s$. The filament's orientation is described by an angle $\theta(s)$ and it can grow in-line ($\epsilon_{\rm g}$) and out-of-line ($\kappa_{\rm g}$). The filament is coupled to a substrate at height $w = 0$ through ``springs'' of stiffness $k$ (see Eq.~\eqref{eq:TotFreeEnergy}), and there is a preferred orientation $\theta_{\rm p}$. (b) Sketch of our framework, in which an elastic filament grows and feedback allows the filament's state (strain $\epsilon$, orientation $\theta$, and curvature $\kappa$) to modify the growth via the feedback laws in Eq.~\eqref{eq:GeneralFeedback}.}
\label{fig:Model}
\end{figure}

\textit{Model}---The filament is assumed to be an elongated quasi-one-dimensional structure embedded in a two-dimensional plane and parametrized by a coordinate $s \in [0, L(t)]$, where $L(t)$ is the total length, which changes over time due to growth. Growth is introduced as an intrinsically driven change in the filament's ground state over time and is represented by an in-line axial displacement field $u(s,t)$ (and a concomitant strain $\epsilon=\partial_s u$) and an out-of-line transverse displacement field $w(s,t)$ (and a concomitant orientation $\theta=\partial_s w$ and curvature $\kappa=\partial_s^2 w$). See Fig.~\ref{fig:Model}(a) for a sketch and SM Tab.~S1 for a list of all parameters. We assume that the filament behaves as an elastic structure with a free energy density that can be expressed as
\begin{equation}
\label{eq:TotFreeEnergy}
\mathcal{F}(t) = \int \text{d}s \left[\, \frac{S}{2} \epsilon_\text{e}(s,t)^2 + \frac{B}{2} \kappa_\text{e}(s,t)^2  + \frac{k}{2} w(s,t)^2\right]\;,
\end{equation}
where $S$ and $B$ are stretching and bending stiffnesses, respectively, and $\epsilon_\text{e}(s,t)=\partial_s u - \epsilon_{\rm g}$ and $\kappa_\text{e}(s,t)=\partial_s^2 w - \kappa_{\rm g}$ are the elastic strain and curvature due to incompatibility with the in-line growth strain $\epsilon_{\rm g}$ and the out-of-line growth curvature $\kappa_{\rm g}$~\cite{efrati2009,liang2009,al-mosleh2023}. See SM Sec.~SI for further details. We further add a Hookean contribution describing the coupling to a substrate, where $k$ is an elastic constant that penalizes transverse deformations $w(s,t)$. We note that our framework is limited to small deformations, where the control framework is most relevant, because the state can then be sensed, and regulatory feedback implemented, more readily. We also note that growth affects the free energy only if it induces strain or curvature; otherwise, it does not appear in this framework.
The linear equations determining the energy-minimizing shape are then found to be (see Fig.~\ref{fig:Model}(b))
\begin{subequations}
\label{eq:EOM}
\begin{align}
&S \partial_s \left[\partial_s u(s,t) - \epsilon_{\rm g}(s,t)\right] = 0 \;, \\
&B \partial^2_s \left[\partial_s^2 w(s,t) - \kappa_{\rm g}(s,t)\right] + k w(s,t) = 0 \;. \label{eq:EOM2}\end{align}
\end{subequations}
We see that if the growth fields were arbitrary, this would result in internal stresses accumulating over time due to geometric incompatibility. The fact that this is not observed in most naturally developing systems suggests that the growth fields are regulated by the state of the system, thereby limiting the uncontrolled accumulation of internal stresses (see Fig.~\ref{fig:Model}(b)). The system's state comprises the strain $\epsilon$, orientation $\theta$ (relative to the substrate or some other environmental cue $\theta_{\rm p}$), and curvature/shape $\kappa$, so that the general leading-order feedback law must have the form:
\begin{subequations}
\label{eq:GeneralFeedback}
\begin{align}
\partial_t \epsilon_{\rm g}(s,t) = &\int \text{d}s' \text{d}t' \left[F_{\epsilon} \epsilon(s',t') + F_{\theta} \sin\left(\theta(s',t')-\theta_\text{p}\right) \right. \nonumber \\
& \left.  + F_{\kappa} \kappa(s',t') \right] + \chi_\epsilon(s,t) \;, \label{eq:GeneralFeedback1} \\
\partial_t \kappa_{\rm g}(s,t) = &\int \text{d}s' \text{d}t' \left[G_{\epsilon} \epsilon(s',t') + G_{\theta} \sin\left(\theta(s',t')-\theta_\text{p}\right) \right. \nonumber \\
& \left.  + G_{\kappa} \kappa(s',t') \right] + \chi_\kappa(s,t)  \;. \label{eq:GeneralFeedback2}
\end{align}
\end{subequations}
Here, $F_i \equiv F_i(t,t',s,s')$ and $G_i \equiv G_i(t,t',s,s')$ are integral kernels that allow the growth fields $\{\epsilon_{\rm g},\kappa_{\rm g}\}$ at coordinates $(s,t)$ to depend on the state fields $i\in\{\epsilon,\theta,\kappa\}$ at other positions $s'$ and earlier times $t'$, corresponding respectively to nonlocal and time-delayed feedback.
The growth will in general fluctuate due to sources that could be internal (e.g., variability in cell growth) or external (e.g., fluctuating environmental conditions). We combine all these effects into a single effective noise $\bm{\chi} = \{\chi_\epsilon, \chi_\kappa\}$, which has zero mean and white-noise correlations encoded by
\begin{subequations}
\label{eq:NoiseCorrelatorMain}
\begin{align}
&\left\langle \chi_\epsilon(s,t)  \chi^\dagger_\epsilon(s',t') \right\rangle = D_u \delta(s - s') \delta(t - t') \\
&\left\langle \chi_\kappa(s,t)  \chi^\dagger_\kappa(s',t') \right\rangle = D_w \delta(s - s') \delta(t - t') \;,
\end{align}
\end{subequations}
with $D_u$ and $D_w$ constants \footnote{While a scale-dependent noise spectrum may be more realistic (see, e.g., S. Armon, M. Moshe, and E. Sharon, \href{https://doi.org/10.1038/s42005-021-00626-z}{Commun. Phys. \textbf{4}, 1 (2021)}), we restrict ourselves to white noise for simplicity.}. Eqs.~\eqref{eq:EOM}--\eqref{eq:NoiseCorrelatorMain} constitute a complete linearized set of equations for $\{\epsilon_{\rm g}, \kappa_{\rm g}, u, w\}$ encoding the mutual coupling of growth and shape. This sets the stage for the investigation of the linear stability of a growing filament that is initially straight with $\theta(s,t = 0) = \theta_\text{p}$.

\textit{Local, instantaneous feedback}---For $k = 0$, corresponding to no coupling to a substrate, Eqs.~\eqref{eq:EOM} and~\eqref{eq:GeneralFeedback} can be combined into
\begin{equation}
\partial_s \partial_t \bm{V} = \int \bm{G} \cdot \bm{V}\, \text{d}s' \text{d}t' + \bm{\chi}(s,t) \;,
\label{eq:CompactEqns}
\end{equation}
where $\bm{V}(s,t) = (u(s,t),\theta(s,t))^\top$, $\bm{G}$ is the matrix of integral kernels, and $\bm{\chi}$ is the vector-valued noise (see SM Sec.~SIII for the full expressions).
First, we analyze the simplest case of local, instantaneous feedback, that is 
\begin{subequations}
\label{eq:LocalFeedback}
\begin{align}
F_i(s,s',t,t') &= - f_i \delta(s - s') \delta(t-t') \;, \\
G_i(s,s',t,t') &= - g_i \delta(s - s') \delta(t-t') \;.
\end{align}
\end{subequations}
The coefficients $f_i$ quantify how each local state variable $i$ regulates axial growth, whereas the coefficients $g_i$ quantify how the same state variables regulate curvature growth. Note that positive values of $f_i,g_i$ correspond to negative feedback: a positive local value of state variable $i$ decreases the corresponding growth rate. Conversely, negative values correspond to positive feedback.
\begin{figure}
\centering
\includegraphics[width=1.\columnwidth]{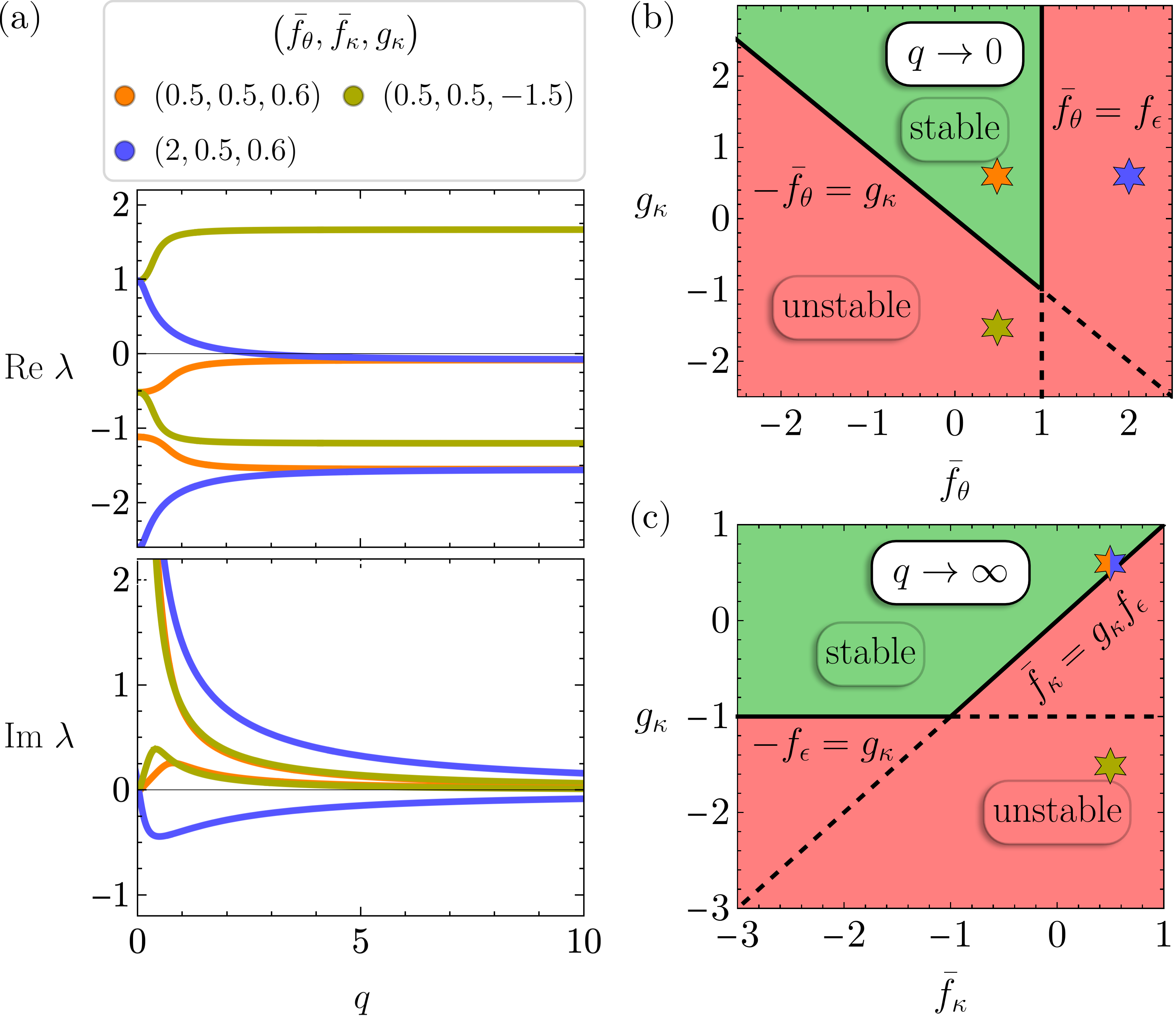}
\caption{\textbf{Local, instantaneous feedback.} (a) The real (top) and imaginary (bottom) parts of the eigenvalues for different parameter values, with $f_{\epsilon} = g_\theta = 1$, obtained from the stability matrix given by Eq.~\eqref{eq:KExpression}. The coefficients $f_i$ and $g_i$ denote feedback from the state variables $i\in\{\epsilon,\theta,\kappa\}$ onto axial and curvature growth, respectively. $\bar{f}_i \propto f_i g_\epsilon$ is the rescaled parameter with ``coupling'' $g_\epsilon$. (b) The stability diagram summarizes the regions of linear stability and instability in the long-wavelength limit ($q \to 0$), with $\bar{f}_\theta$ and $g_{\kappa}$ as the relevant parameters. Stable and unstable regions are shown in green and red, respectively, and stars indicate the parameter values used in panel (a). (c) The corresponding diagram for the short-wavelength limit ($q \to \infty$), where the relevant parameters are $\bar{f}_\kappa$ and $g_{\kappa}$. Throughout, we set $f_{\epsilon} = 1$.
}
\label{fig:LocalFeedback}
\end{figure}

In the deterministic limit, we perform a linear stability analysis using the ansatz $(u,\theta) = (u_0,\theta_0) e^{i q s + \lambda t}$, which allows us to write Eq.~\eqref{eq:CompactEqns} as $\left[\lambda \mathbf{1} - \bm{K} \right] \cdot \bm{V} = 0$ with
\begin{equation}
\label{eq:KExpression}
\bm{K}(q) = \begin{pmatrix} 
    - f_{\epsilon} & -f_{\kappa}+i f_{\theta}/q    \\ 
    - g_{\epsilon} & - g_{\kappa} + i g_{\theta}/q 
    \end{pmatrix} \;,
\end{equation}
and we solve $\det\left[\lambda \mathbf{1} - \bm{K} \right] = 0$ for the growth rates $\lambda$. Defining $\bar{f}_\theta \equiv f_{\theta} g_{\epsilon}/g_{\theta}$ and $\bar{f}_\kappa \equiv f_{\kappa} g_{\epsilon}$ removes the explicit dependence on $g_{\epsilon}$ from what follows. Thus, the rescaled parameters measure the effective strength of axial-growth feedback after weighting by $g_\epsilon$, the coupling from strain to curvature growth.
Fig.~\ref{fig:LocalFeedback}(a) shows the real and imaginary parts of the eigenvalues $\lambda(q)$ for different parameter values. In the long-wavelength limit ($q \to 0$), $\text{Im }\lambda \sim g_{\theta}/q$ and thus diverges whenever $g_{\theta} \neq 0$, indicating oscillatory filament dynamics. These oscillations are damped only if $\text{Re }\lambda < 0$, which is the condition for stability. The resulting stability diagram and corresponding bounds are shown in Fig.~\ref{fig:LocalFeedback}(b). In the short-wavelength limit ($q \to \infty$), the imaginary part approaches a constant, whereas the real part is again negative only over a restricted region of parameter space; see Fig.~\ref{fig:LocalFeedback}(c). Since the real parts vary monotonically with $q$, stability in both asymptotic limits guarantees stability at all wavelengths. In the special case of equal coefficients, $f_i = g_i = g$, the condition $\text{Re } \lambda < 0$ requires $g>0$, so the feedback is purely damping. More generally, however, the parameters are not constrained to be positive; see Fig.~\ref{fig:LocalFeedback}. For example, sufficiently strong proprioceptive feedback $g_\kappa$ stabilizes the system independently of the remaining feedback parameters (additional details are provided in SM Sec.~SIII).

We now consider the effect of noise. States that are stable in the deterministic limit may be destabilized by noise. To investigate the stability in the presence of noise, we consider the angular fluctuations, which vanish for a straight filament. With $\hat \theta(q,\Omega)$ the Fourier transform of $\theta(s,t)$, where $\lambda = i \Omega$, we have
\begin{align}
\label{eq:NoiseCorr}
&\left\langle\theta(s,t)^2\right\rangle \simeq \int \frac{\text{d}q\, \text{d}\Omega\, \text{d}q'\, \text{d}\Omega'}{\left(2 \pi\right)^4} \left\langle \hat \theta(q,\Omega) \hat \theta^\dagger(q',\Omega')\right\rangle \;.
\end{align}
We leave the derivation of the explicit expression to the Appendix and SM Sec.~SIII, and present here a simplified derivation that preserves the essential results while shortening the expressions and clarifying their origin. Specifically, we set $f_i = 0$ and $g_\epsilon = 0$. Combining Eqs.~\eqref{eq:CompactEqns} and~\eqref{eq:LocalFeedback}, and Fourier transforming, we obtain a single equation for the angle field, independent of $u$:
\begin{equation}
\label{eq:ReducedAngleEq}
q  \Omega \hat \theta = g_{\theta} \hat \theta + i q g_{\kappa} \hat \theta - \hat\chi_\kappa \;.
\end{equation}
This can be solved straightforwardly for $\hat \theta$. Using Eq.~\eqref{eq:NoiseCorrelatorMain}, we then find that Eq.~\eqref{eq:NoiseCorr} can be written as
\begin{subequations}
\begin{align}
\label{eq:AngleFluctuationsOrig}
&\left\langle\theta(s,t)^2\right\rangle \simeq \int \frac{\text{d}q\, \text{d}\Omega}{\left(2 \pi\right)^2} P_\theta(q,\Omega) \;,
\end{align}
where
\begin{equation}
\label{eq:LocalPTheta}
P_\theta(q,\Omega) = \frac{D_w}{q^2 g_\kappa^2 + (g_\theta - q \Omega)^2} \;.
\end{equation}
\end{subequations}
Thus, $P_\theta \to 0$ as $q \to \infty$, and noise is irrelevant at short length scales. At large scales, however, stability requires $P_\theta \sim q^k$ with $k \ge 0$, since the correlation function in Eq.~\eqref{eq:AngleFluctuationsOrig} scales as $\int \text{d}q\, P_\theta$. That is, without feedback ($g_i = 0$), $P_{\theta} \sim q^{-2}$, and height fluctuations always diverge for small $q$ if noise is present ($D_w \neq 0$). This divergence persists even in the presence of proprioceptive feedback ($g_{\kappa} \neq 0$), because bending becomes negligible as $q \to 0$, since the corresponding term $\sim \partial_s^2 w$ is small, as is evident from Eq.~\eqref{eq:ReducedAngleEq}. Consequently, proprioceptive feedback cannot protect against long-wavelength instabilities, which involve small-curvature deformations. By contrast, if $g_{\theta} \neq 0$, $P_\theta$ remains finite and $P_\theta \sim D_w/g_{\theta}^2$ as $q \to 0$, because the orientation term $\sim \partial_s w$ dominates over the bending term $\sim \partial_s^2 w$ for small $q$ (cf. Eq.~\eqref{eq:ReducedAngleEq}). Hence, because orientation sensing acts along the entire filament, it can suppress long-wavelength deviations from the straight state. The same conclusion holds for the full problem: with local feedback, orientation sensing is necessary to prevent noise-induced instabilities (see Appendix and SM Sec.~SIII).

\textit{Nonlocal, time-delayed feedback}---In many biological systems, sensing and response are mediated by information transport, for example through hormonal signaling, thereby producing a nonlocal response to the system's state. We capture this by modifying the local feedback kernels in Eq.~\eqref{eq:LocalFeedback} to include a nonlocal contribution of the form
\begin{subequations}
\label{eq:FTGreen}
\begin{align}
F_i(s,t) &= - f_i \delta(s) \delta(t) - \varphi_i G_{\rm S}(s,t) \;, \\
G_i(s,t) &= - g_i \delta(s) \delta(t) - \gamma_i G_{\rm S}(s,t) \;.
\end{align}
\end{subequations}
The coefficients $\varphi_i$ and $\gamma_i$ set the strength of the nonlocal feedback mediated by the transported signal $G_{\rm S}(s,t)$. To describe signal propagation, we follow Ref.~\cite{al-mosleh2023} and consider the simple case of diffusive transport with a characteristic delay timescale $\Gamma^{-1}$. Assuming that elastic relaxation is much faster than signal propagation, we work in the elastostatic limit, for which $\Delta G_{\rm S}(s,t) = -\delta(s) e^{-\Gamma t}$. The Fourier transforms of the feedback kernels in Eq.~\eqref{eq:FTGreen} then take the form
\begin{subequations}
\label{eq:FourierGreenNonLocal}
\begin{align}
\hat{F}_i(q,\Omega) &= - f_i - \frac{\varphi_i}{\Gamma + i \Omega} \frac{1}{q^2} \;, \\
\hat{G}_i(q,\Omega) &= - g_i - \frac{\gamma_i}{\Gamma + i \Omega} \frac{1}{q^2} \;.
\end{align}
\end{subequations}
\begin{figure}
\centering
\includegraphics[width=1.\columnwidth]{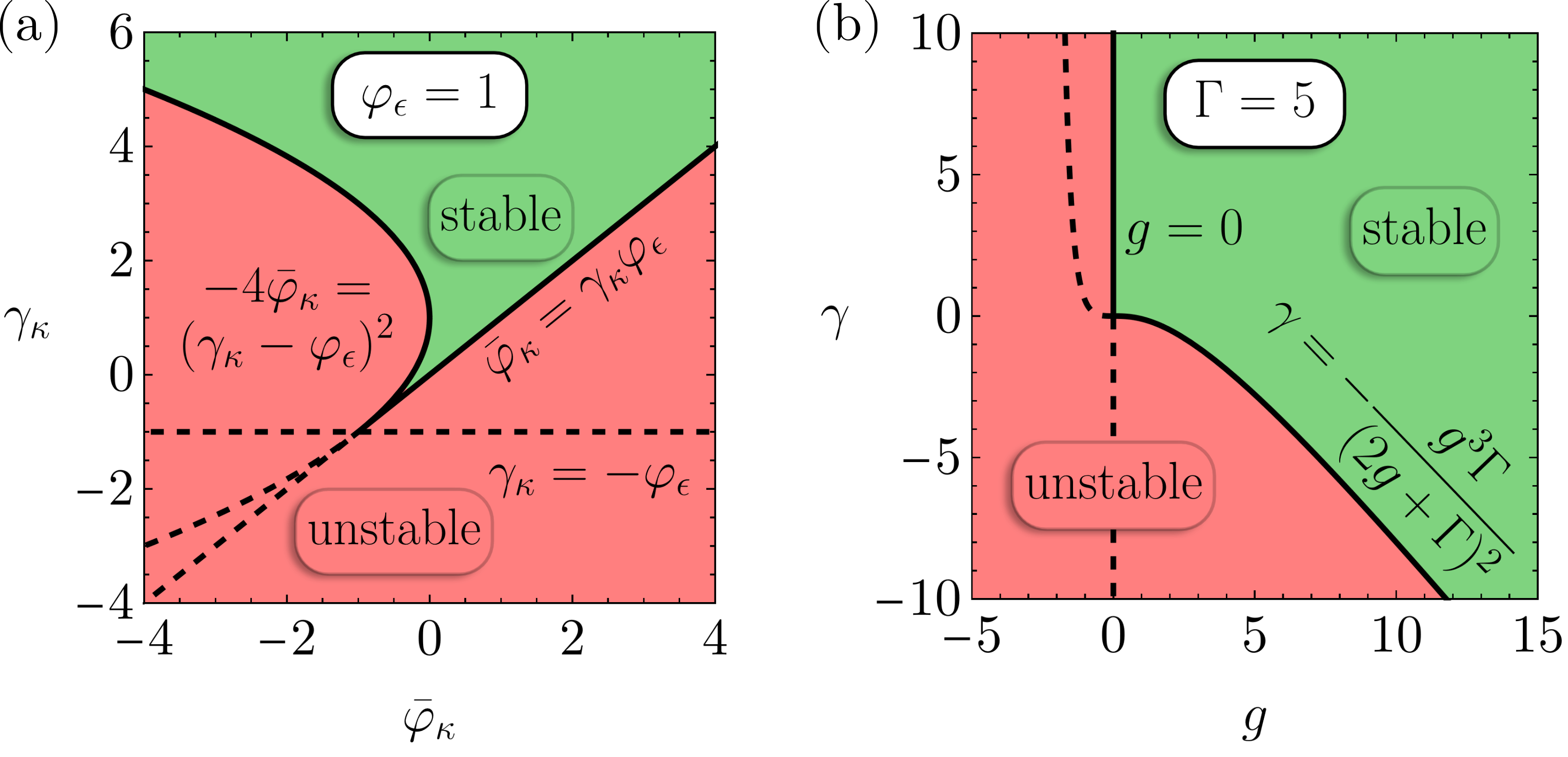}
\caption{\textbf{Nonlocal feedback.} (a) Stability diagram for nonlocal feedback in the limit where local feedback is negligible ($f_i = g_i = 0$), with $\bar{\varphi}_\kappa = \varphi_\kappa \gamma_\epsilon$ rescaled by the ``coupling'' $\gamma_\epsilon$. We choose $\varphi_\epsilon = 1$ here; see SM Sec.~SIV for a different choice. The axes $\gamma_\kappa$ and $\bar{\varphi}_\kappa$ denote the strengths of the nonlocal feedback by which the system's curvature regulates the growth fields. (b) Stability diagram assuming equal magnitudes for the feedback coefficients, i.e., $f_i = g_i = g$ and $\varphi_i = \gamma_i = \gamma$. We choose $\Gamma = 5$ here; see SM Sec.~SIV for a different choice and for the derivation of the bounds in (a) and (b). We used $\gamma_\theta = 0$ and $\bar{\varphi}_\theta = 0$ throughout.}
\label{fig:NonLocalFeedback}
\end{figure}
In the absence of noise, we can again analyze the linear stability by solving the characteristic equation $\det\left[\lambda \mathbf{1} - \bm{K}(q, \lambda)\right] = 0$ for $\lambda$, as above, where $\bm{K}$ is given by Eq.~\eqref{eq:KExpression} after the replacements $f_i \to \hat{F}_i$ and $g_i \to \hat{G}_i$. 
There are now four eigenvalues (see SM Sec.~SIV for their expressions). As $q \to \infty$, the parameters $\varphi_i$ and $\gamma_i$ do not enter, and we find the same conditions as above for the real parts to be nonpositive, together with the condition $\Gamma > 0$. On the other hand, for $q \to 0$, the nonlocal feedback is relevant due to the $q^{-2}$ scaling in Eq.~\eqref{eq:FourierGreenNonLocal}. 
We find that taking the limits $\gamma_{\theta},\, \varphi_{\theta} \gamma_{\epsilon} \to 0$ is necessary for stability, as otherwise $\text{Re } \lambda$ always diverges (see SM Sec.~SIV). After taking these limits, the expressions for $\lambda$ are too complex to write explicitly, and thus we consider two simplifying cases: dominant nonlocal feedback ($f_i, g_i \sim 0$) and equal feedback magnitudes ($f_i = g_i = g$ and $\varphi_i = \gamma_i = \gamma$). It is then straightforward to analyze the eigenvalues as before. The resulting stability diagrams and bounds are shown in Fig.~\ref{fig:NonLocalFeedback}; see SM Sec.~SIV for the explicit derivation.

In the presence of noise, we compute the angular correlation function as before. The full discussion is presented in the Appendix and SM Sec.~SIV. Here we again focus on a simple case that highlights the main features. Specifically, we restrict the analysis to the height equation and set $F_i = 0$ and $G_\epsilon = 0$. Analogously to Eq.~\eqref{eq:ReducedAngleEq}, we then find
\begin{equation}
\label{eq:ReducedHeightEqNonLoc}
q \Omega \hat \theta = - \hat{G}_{\theta}(q,\Omega) \hat \theta - i q \hat{G}_{\kappa}(q,\Omega) \hat \theta - \hat\chi_\kappa \;.
\end{equation}
As in the local case, the effect of feedback is relevant only in the limit $q \to 0$. Solving this equation for $\hat \theta$ and computing the correlation function, we obtain
\begin{equation}
P_{\theta} = \frac{D_w \left( \Gamma^2 + \Omega^2\right) q^4}{\gamma_\theta^2 + (2\Gamma g_\theta \gamma_\theta + \gamma_\kappa^2) q^2 + \mathcal{O}\left(q^{3}\right)} \;,
\label{eq:PThetaNonLocal}
\end{equation}
where we have expanded the denominator to highlight the role of nonlocal feedback for $q \to 0$. As expected from the scaling in Eqs.~\eqref{eq:FourierGreenNonLocal}, the nonlocal terms $\propto \gamma_i$ dominate over the local terms $g_i$. Moreover, this expression is structurally similar to Eq.~\eqref{eq:LocalPTheta}, but because of the $q^{-2}$ scaling of the nonlocal feedback, the proprioceptive term alone is now sufficient to stabilize the system: $P_{\theta} \to 0$ as $q \to 0$ even in the absence of orientation sensing ($G_\theta = 0$). In addition, we find several other minimal combinations of feedback parameters for which $P_\theta$ remains finite, with the scaling of $P_\theta(q)$ depending on which parameters are nonzero. These cases are summarized in Tab.~\ref{tab:tab1} in the Appendix. Hence, appropriate combinations of nonlocal feedback can suppress noise-induced long-wavelength instabilities, while short wavelengths remain unaffected.

\textit{Elastic substrate}---Finally, we consider the effect of a substrate to which the growing filament is coupled elastically. We thus take $k \neq 0$ such that translational symmetry is now broken. Furthermore, for simplicity, we set $f_{\epsilon} = g_{\kappa} = g$ and all other feedback parameters are set to zero, to focus on the effect of the elastic substrate. In the deterministic limit, we then have $\bm{K} = \text{diag}[-g, -g q^4/(q^4 + \tilde{k})]$, with $\tilde{k} = k/B$, from which the eigenvalues are found immediately. Hence, as before, the flat state is stable if $g > 0$. The elastic substrate does not modify the stability. However, in the presence of noise, we obtain from Eq.~\eqref{eq:EOM2} that
\begin{equation}
i \Omega q^4 \hat{w} + g q^4 \hat{w} + q^2 \hat{\chi}_\kappa + i \Omega \tilde{k} \hat{w} = 0
\end{equation}
and it follows that the substrate term is dominant as $q \to 0$. In this limit, $\hat{w} \sim q^2 \hat{\chi}_\kappa / \tilde{k}$, so that $P_\theta \sim q^6$.
The full correlation function is found to be
\begin{equation}
P_{\theta} = \frac{D_w q^6}{g^2 q^8 + (\tilde{k} + q^4)^2 \Omega^2} \underset{q \to 0}{\sim} \frac{D_w q^6}{\tilde{k}^2 \Omega^2} + \mathcal{O}\left(q^7\right) \;.
\end{equation}
Hence, coupling to an elastic substrate can strongly suppress noise-induced long-wavelength fluctuations, while short length scales are not affected by the presence of the elastic substrate.

\textit{Discussion}---We have investigated the stabilizing effect of feedback in a stochastically growing elastic filament with strain, orientation, and curvature sensing. In the deterministic limit, we find constraints on the feedback that must be satisfied to guarantee linear stability. In particular, a system with nonlocal orientation sensing is always unstable. In the presence of noise, orientation sensing is required for stability under local feedback, while under nonlocal feedback, proprioception can also suppress divergences. Different combinations of parameters are found to give rise to different scalings of the angle-angle correlation function, as summarized in Tab.~\ref{tab:tab1}. In general, orientation sensing suppresses divergences more strongly than the corresponding curvature sensing, and large-scale deviations from the flat state are most strongly suppressed through coupling to an elastic substrate.
While much is known about growth and feedback, the effects of noise remain less well understood. In a few cases, such as plant growth, it has been shown that gravity/orientation sensing is required for straight growth~\cite{okamoto2015,berut2018,moulia2021}. 
However, as noted in the introduction, similar feedback mechanisms operate across organisms and scales, yet the role of noise has rarely been addressed. Our results suggest experimental strategies, such as measuring angle fluctuations, to identify which feedback mechanisms are present and which are required for stable growth.

Our analysis of the growth of a one-dimensional filament in two-dimensional space excludes geometric torsion and physical twist. For small three-dimensional deformations of a helically growing filament, many of our results can be easily generalized by mapping $\kappa \to \kappa(\sin\Phi(s),\cos\Phi(s))$. Comparing our results with those for the growth of sheets, we note that our results for purely proprioceptive feedback are consistent with those of Ref.~\cite{al-mosleh2023}. However, the results for orientation sensing probe a different modality, and it remains unclear how this type of sensing extends to sheets, since their orientation cannot be described by a single scalar $\theta$, but instead requires a vector field corresponding to the local normal. On the other hand, substrate coupling is expected to have a similar stabilizing effect and may be relevant to structurally colored petal patterns~\cite{antonioukourounioti2013}, where the impact of noise has not yet been characterized.

We note that our exploratory study is limited to the linear regime. Natural next steps include the study of feedback-driven growth in highly curved and twisted filaments and sheets. Additionally, because both growth and sensing are generally noisy, accounting for measurement errors and characterizing the role of self-estimation and shape filtering in growing structures are important goals for future work.

\section*{Acknowledgments}
This work was partly supported by the NWO Rubicon grant (L.A.H.), the Simons Foundation (L.M.) and the Henri Seydoux Fund (L.M.).

\section*{Appendix}
Here we present the full derivation of the explicit expression for the noise-induced angular fluctuations $\left\langle\theta(s,t)^2\right\rangle$. We have
\begin{equation}
\label{eq:AngleFluctuationsOrigSI}
\left\langle\theta(s,t)^2\right\rangle \simeq \int \frac{\text{d}q\, \text{d}\Omega}{\left(2 \pi\right)^2} \int \frac{\text{d}q'\, \text{d}\Omega'}{\left(2 \pi\right)^2} \left\langle \hat \theta(q,\Omega) \hat \theta^\dagger(q',\Omega')\right\rangle \;,
\end{equation}
where $\hat \theta(q,\Omega)$ is the Fourier transform of $\theta(s,t)$.
We thus need to find an expression for the correlation function appearing here. To this end we assume that the noise fields introduced in the growth equations have the white noise spectrum in Eq.~\eqref{eq:NoiseCorrelatorMain}. Furthermore, these noise terms are related to the noise fields associated with $u$ and $\theta$ through $\chi_\epsilon(s,t) = \partial_s \chi_u(s,t)$ and $\chi_\kappa(s,t) = \partial_s \chi_\theta(s,t)$ such that in Fourier space
\begin{equation}
\label{eq:CorrNoiseFin}
\left\langle\hat{\bm{X}}(q,\Omega) \hat{\bm{X}}^\dagger(q',\Omega') \right\rangle = \left(2\pi\right)^2 \bm{D}(q) \delta(q - q') \delta(\Omega - \Omega') \;,
\end{equation}
where $\bm{D}(q) = \text{diag}(D_u/q^2,D_w/q^2)$ and $\hat{\bm{X}} = (\hat{\chi}_u(q,\Omega), \hat{\chi}_\theta(q,\Omega))^\top$.
We can therefore write the local and instantaneous limit of Eqs.~\eqref{eq:CompactEqns} in Fourier space as
\begin{equation}
\label{eq:NoisyEquationGeneral}
i \Omega \mathbf{1} \cdot \hat{\bm{V}} = \bm{K}(q) \cdot \hat{\bm{V}}  + \hat{\bm{X}} \;, 
\end{equation}
with $\hat{\bm{V}}  = (\hat u(q,\Omega), \hat \theta(q,\Omega))^\top$ and $\bm{K}(q)$ as before (cf. Eq.~\eqref{eq:KExpression}).
Solving for $\hat{\bm{V}}$ then yields
\begin{equation}
\hat{\bm{V}} = \left[i \Omega \mathbf{1} - \bm{K}(q) \right]^{-1} \cdot \hat{\bm{X}} \equiv \bm{T} \cdot \hat{\bm{X}} \;,
\end{equation}
where we defined $\bm{T} \equiv \left[i \Omega \mathbf{1} - \bm{K}(q) \right]^{-1}$ to make the following expressions shorter.
From this the angular correlation function is found. Namely, using Eq.~\eqref{eq:CorrNoiseFin}, we find from the previous equation that
\begin{subequations}
\begin{equation}
\left\langle \hat{\bm{V}}(q,\Omega) \hat{\bm{V}}^\dagger(q',\Omega') \right\rangle = \bm{P}(q, q', \Omega, \Omega') \delta(q - q') \delta(\Omega - \Omega') \;,
\end{equation}
where we have defined the correlation matrix
\begin{equation}
\label{eq:PDef}
\bm{P}(q, q', \Omega, \Omega') \equiv \bm{T}(q,\Omega) \cdot \bm{D}(q) \cdot \bm{T}^\dagger(q',\Omega') \;.
\end{equation}
\end{subequations}
The $(2,2)$-component of $\bm{P}$ then yields the correlation function of the angle field. Substituting the expression into Eq.~\eqref{eq:AngleFluctuationsOrigSI}, we find
\begin{equation}
\left\langle\theta(s,t)^2\right\rangle \simeq \int \frac{\text{d}q\, \text{d}\Omega}{\left(2 \pi\right)^2}\ P_{22}(q,\Omega)  \;,
\end{equation}
where $P_{22}(q,\Omega) = T_{2i} D_{ij} T_{j2}^\dagger$. Using the definitions of $\bm{T}$ and $\bm{D}$, we then find
\begin{subequations}
\begin{equation}
P_\theta \equiv P_{22} = \frac{D_u g_{\epsilon}^2 + D_w \left[f_{\epsilon}^2 +\Omega^2 \right]}{|\mathcal{D}|^2} \;,
\end{equation}
where
\begin{align}
\label{eq:DEq}
\mathcal{D} &= q \left[f_{\kappa} g_{\epsilon} - f_{\epsilon} g_{\kappa} - i (f_{\epsilon} + g_{\kappa}) \Omega + \Omega^2 \right] \nonumber \\
& + i\left[f_{\epsilon} g_{\theta} - f_{\theta} g_{\epsilon} + i g_{\theta} \Omega \right] \;.
\end{align} 
\end{subequations}
Thus, $P_\theta \to 0$ as $q \to \infty$, so noise does not affect short length scales. At large scales, however, without feedback ($f_i, g_i = 0$), $P_{\theta} \sim q^{-2}$. This divergence remains even when $f_{\epsilon}$, $f_{\theta}$, $f_{\kappa}$, and $g_{\kappa}$ are nonzero, in which case
\begin{equation}
P_{\theta} = \frac{D_w}{q^2 \left(g_{\kappa}^2 + \Omega^2\right)}
\label{eq:PThetaDiv}
\end{equation}
and height fluctuations always diverge for small $q$ if $D_w \neq 0$. By contrast, if $g_{\theta} \neq 0$ or $\bar{f}_\theta \neq 0$, the system is stabilized. We find
\begin{equation}
\label{eq:P22Theta}
\left.P_{\theta}\right|_{q \to 0} = \frac{D_u g_{\epsilon}^2 + D_w \left(f_{\epsilon}^2 + \Omega^2\right)}{\left(f_\theta g_\epsilon - g_{\theta} f_{\epsilon} \right)^2 + g_{\theta}^2 \Omega^2} + \mathcal{O}(q) \;.
\end{equation}
The strain-orientation term $f_\theta$ only appears coupled to $g_\epsilon$; $g_\theta$ appears on its own.

For nonlocal feedback, expanding about $q = 0$ yields
\begin{equation}
\left.P_{\theta}\right|_{q \to 0} = \frac{D_u \gamma_{\epsilon}^2 + D_w \varphi_{\epsilon}^2}{\left(\varphi_{\theta} \gamma_\epsilon - \gamma_{\theta} \varphi_{\epsilon}\right)^2} \left( \Gamma^2 + \Omega^2\right) q^4 + \mathcal{O}\left(q^{6}\right) \;,
\label{eq:PThetaFour}
\end{equation} 
which is similar to Eq.~\eqref{eq:P22Theta} but with a different scaling. When taking $\gamma_{\theta}, \varphi_\theta \gamma_\epsilon \to 0$ (to guarantee stability in the deterministic limit) we instead find
\begin{equation}
\left.P_{\theta}\right|_{q \to 0} = \frac{D_u \gamma_{\epsilon}^2 + D_w \varphi_{\epsilon}^2}{\left(\varphi_\kappa \gamma_\epsilon - \gamma_{\kappa} \varphi_{\epsilon} \right)^2} \left( \Gamma^2 + \Omega^2\right) q^2 + \mathcal{O}\left(q^{4}\right) \;,
\end{equation}
i.e., proprioceptive terms ``replace'' the orientation-sensing terms in Eq.~\eqref{eq:PThetaFour}.

\renewcommand{\arraystretch}{1.3}
\begin{table}
\begin{tabular}{| c | | c |}
\hline
Scaling & Parameters \\
\hline
$P_{\theta} \sim q^{0}$ & $g_{\theta} \neq 0$ \; or \; $f_{\theta} \gamma_{\epsilon} \neq 0$ \; or \; $f_{\theta} g_{\epsilon} \neq 0$  \\ 
\hline
$P_{\theta} \sim q^{2}$ & $\gamma_{\kappa} \neq 0$ \; or \; $\varphi_{\kappa} \gamma_{\epsilon} \neq 0$ \; or \; $\varphi_{\kappa} g_{\epsilon} \neq 0$  \\
\hline
$P_{\theta} \sim q^{4}$ & $\gamma_{\theta} \neq 0$ \; or \; $\varphi_{\theta} \gamma_{\epsilon} \neq 0$ \; or \; $\varphi_{\theta} g_{\epsilon} \neq 0$ \\
\hline
$P_{\theta} \sim q^{6}$ & $k \neq 0$ (elastic substrate) \\
\hline
\end{tabular}
\caption{\textbf{Scaling of angle fluctuations.} Left: scalings for which $P_\theta$ (angular correlation) remains finite as $q \to 0$. Right: corresponding nonzero parameters (all others set to zero), with $f_i,g_i$ and $\gamma_i,\varphi_i$ local and nonlocal feedback parameters, respectively. As explained in the main text, the deterministic system is unstable if $\gamma_{\theta} \neq 0$ or $\varphi_{\theta} \gamma_{\epsilon} \neq 0$.}
\label{tab:tab1}
\end{table}

As explained in the main text, having $\gamma_\kappa \neq 0$ while all other parameters vanish is sufficient to obtain a finite $P_\theta(q)$ as $q \to 0$. In addition, we find several other cases for which $P_\theta$ is finite. All minimal combinations of parameters are summarized in Tab.~\ref{tab:tab1} together with the respective $q$-scaling of $P_\theta(q)$ as $q \to 0$. Note that the curvature-strain terms $\{g_\epsilon, \gamma_\epsilon\}$ only appear coupled to either the strain-orientation terms $\{f_\theta, \varphi_\theta\}$ or the strain-curvature terms $\{f_\kappa, \varphi_\kappa\}$. Different combinations give rise to different scalings of $P_\theta(q)$. By contrast, $\{f_i, \varphi_i\}$ never appear independently, whereas $\{g_\theta, \gamma_\theta, \gamma_\kappa\}$ do. The full expressions are given in SM Sec.~SIV.

\bibliography{Biblio.bib}

@article{al-mosleh2023,
  title = {How to {{Grow}} a {{Flat Leaf}}},
  author = {{al-Mosleh}, Salem and Mahadevan, L.},
  year = {2023},
  journal = {Phys. Rev. Lett.},
  volume = {131},
  number = {9},
  pages = {098401},
  doi = {10.1103/PhysRevLett.131.098401}
}

@article{antonioukourounioti2013,
  title = {Buckling as an Origin of Ordered Cuticular Patterns in Flower Petals},
  author = {Antoniou Kourounioti, Rea L. and Band, Leah R. and Fozard, John A. and Hampstead, Anthony and Lovrics, Anna and Moyroud, Edwige and Vignolini, Silvia and King, John R. and Jensen, Oliver E. and Glover, Beverley J.},
  year = {2013},
  journal = {J. R. Soc. Interface},
  volume = {10},
  number = {80},
  pages = {20120847},
  doi = {10.1098/rsif.2012.0847}
}

@article{armon2021,
  title = {The Multiscale Nature of Leaf Growth Fields},
  author = {Armon, Shahaf and Moshe, Michael and Sharon, Eran},
  year = {2021},
  journal = {Commun. Phys.},
  volume = {4},
  number = {1},
  pages = {1--7},
  doi = {10.1038/s42005-021-00626-z}
}

@article{bastien2013,
  title = {Unifying Model of Shoot Gravitropism Reveals Proprioception as a Central Feature of Posture Control in Plants},
  author = {Bastien, Renaud and Bohr, Tomas and Moulia, Bruno and Douady, St{\'e}phane},
  year = {2013},
  journal = {Proc. Natl. Acad. Sci. U.S.A.},
  volume = {110},
  number = {2},
  pages = {755--760},
  doi = {10.1073/pnas.1214301109}
}

@article{berut2018,
  title = {Gravisensors in Plant Cells Behave like an Active Granular Liquid},
  author = {B{\'e}rut, Antoine and Chauvet, Hugo and Legu{\'e}, Val{\'e}rie and Moulia, Bruno and Pouliquen, Olivier and Forterre, Yo{\"e}l},
  year = {2018},
  journal = {Proc. Natl. Acad. Sci. U.S.A.},
  volume = {115},
  number = {20},
  pages = {5123--5128},
  doi = {10.1073/pnas.1801895115}
}

@article{chelakkot2017,
  title = {On the Growth and Form of Shoots},
  author = {Chelakkot, Raghunath and Mahadevan, L.},
  year = {2017},
  journal = {J. R. Soc. Interface},
  volume = {14},
  number = {128},
  pages = {20170001},
  doi = {10.1098/rsif.2017.0001}
}

@article{damavandi2019,
  title = {Statistics of Noisy Growth with Mechanical Feedback in Elastic Tissues},
  author = {Damavandi, Ojan Khatib and Lubensky, David K.},
  year = {2019},
  journal = {Proc. Natl. Acad. Sci. U.S.A.},
  volume = {116},
  number = {12},
  pages = {5350--5355},
  doi = {10.1073/pnas.1816100116}
}

@article{efrati2009,
  title = {Elastic Theory of Unconstrained Non-{{Euclidean}} Plates},
  author = {Efrati, E. and Sharon, E. and Kupferman, R.},
  year = {2009},
  journal = {J. Mech. Phys. Solids},
  volume = {57},
  number = {4},
  pages = {762--775},
  doi = {10.1016/j.jmps.2008.12.004}
}

@article{hamant2019,
  title = {Are Microtubules Tension Sensors?},
  author = {Hamant, Olivier and Inoue, Daisuke and Bouchez, David and Dumais, Jacques and Mjolsness, Eric},
  year = {2019},
  journal = {Nat. Commun.},
  volume = {10},
  number = {1},
  pages = {2360},
  doi = {10.1038/s41467-019-10207-y}
}

@article{hong2018,
  title = {Heterogeneity and {{Robustness}} in {{Plant Morphogenesis}}: {{From Cells}} to {{Organs}}},
  author = {Hong, Lilan and Dumond, Mathilde and Zhu, Mingyuan and Tsugawa, Satoru and Li, Chun-Biu and Boudaoud, Arezki and Hamant, Olivier and Roeder, Adrienne H. K.},
  year = {2018},
  journal = {Annu. Rev. Plant Biol.},
  volume = {69},
  number = {1},
  pages = {469--495},
  doi = {10.1146/annurev-arplant-042817-040517}
}

@article{jonsson2022,
  title = {Multiple Mechanisms behind Plant Bending},
  author = {Jonsson, Kristoffer and Ma, Yuan and {Routier-Kierzkowska}, Anne-Lise and Bhalerao, Rishikesh P.},
  year = {2022},
  journal = {Nat. Plants},
  volume = {9},
  number = {1},
  pages = {13--21},
  doi = {10.1038/s41477-022-01310-y}
}

@article{liang2009,
  title = {The Shape of a Long Leaf},
  author = {Liang, Haiyi and Mahadevan, L.},
  year = {2009},
  journal = {Proc. Natl. Acad. Sci. U.S.A.},
  volume = {106},
  number = {52},
  pages = {22049--22054},
  doi = {10.1073/pnas.0911954106}
}

@article{meroz2019,
  title = {Spatio-Temporal Integration in Plant Tropisms},
  author = {Meroz, Yasmine and Bastien, Renaud and Mahadevan, L.},
  year = {2019},
  journal = {J. R. Soc. Interface},
  volume = {16},
  number = {154},
  pages = {20190038},
  doi = {10.1098/rsif.2019.0038}
}

@article{moulia2021,
  title = {Fluctuations Shape Plants through Proprioception},
  author = {Moulia, Bruno and Douady, St{\'e}phane and Hamant, Olivier},
  year = {2021},
  journal = {Science},
  volume = {372},
  number = {6540},
  pages = {eabc6868},
  doi = {10.1126/science.abc6868}
}

@article{nakayama2012,
  title = {Mechanical {{Regulation}} of {{Auxin-Mediated Growth}}},
  author = {Nakayama, Naomi and Smith, Richard S. and Mandel, Therese and Robinson, Sarah and Kimura, Seisuke and Boudaoud, Arezki and Kuhlemeier, Cris},
  year = {2012},
  journal = {Curr. Biol.},
  volume = {22},
  number = {16},
  pages = {1468--1476},
  doi = {10.1016/j.cub.2012.06.050}
}

@article{okamoto2015,
  title = {Regulation of Organ Straightening and Plant Posture by an Actin--Myosin {{XI}} Cytoskeleton},
  author = {Okamoto, Keishi and Ueda, Haruko and Shimada, Tomoo and Tamura, Kentaro and Kato, Takehide and Tasaka, Masao and Morita, Miyo Terao and {Hara-Nishimura}, Ikuko},
  year = {2015},
  journal = {Nat. Plants},
  volume = {1},
  number = {4},
  pages = {1--7},
  doi = {10.1038/nplants.2015.31}
}

@article{shraiman2005,
  title = {Mechanical Feedback as a Possible Regulator of Tissue Growth},
  author = {Shraiman, Boris I.},
  year = {2005},
  journal = {Proc. Natl. Acad. Sci. U.S.A.},
  volume = {102},
  number = {9},
  pages = {3318--3323},
  doi = {10.1073/pnas.0404782102}
}

@article{xu2024,
  title = {A 3-Component Module Maintains Sepal Flatness in {{Arabidopsis}}},
  author = {Xu, Shouling and He, Xi and Trinh, Duy-Chi and Zhang, Xinyu and Wu, Xiaojiang and Qiu, Dengying and Zhou, Ming and Xiang, Dan and Roeder, Adrienne H. K. and Hamant, Olivier and Hong, Lilan},
  year = {2024},
  journal = {Curr. Biol.},
  volume = {34},
  number = {17},
  pages = {4007--4020.e4},
  doi = {10.1016/j.cub.2024.07.066}
}

@article{troutwine2020,
  title = {The {{Reissner Fiber Is Highly Dynamic In Vivo}} and {{Controls Morphogenesis}} of the {{Spine}}},
  author = {Troutwine, Benjamin R. and Gontarz, Paul and Konjikusic, Mia J. and Minowa, Ryoko and {Monstad-Rios}, Adrian and Sepich, Diane S. and Kwon, Ronald Y. and {Solnica-Krezel}, Lilianna and Gray, Ryan S.},
  year = {2020},
  journal = {Curr. Biol.},
  volume = {30},
  number = {12},
  pages = {2353--2362.e3},
  doi = {10.1016/j.cub.2020.04.015}
}

@article{wyart2023,
  title = {Cerebrospinal Fluid-Contacting Neurons: Multimodal Cells with Diverse Roles in the {{CNS}}},
  author = {Wyart, Claire and {Carbo-Tano}, Martin and {Cantaut-Belarif}, Yasmine and {Orts-Del'Immagine}, Adeline and B{\"o}hm, Urs L.},
  year = {2023},
  journal = {Nat. Rev. Neurosci.},
  volume = {24},
  number = {9},
  pages = {540--556},
  doi = {10.1038/s41583-023-00723-8}
}

@article{fang2023,
  title = {Enabled Primarily Controls Filopodial Morphology, Not Actin Organization, in the {{TSM1}} Growth Cone in {{{\emph{Drosophila}}}}},
  author = {Fang, Hsiao Yu and Forghani, Rameen and Clarke, Akanni and McQueen, Philip G. and Chandrasekaran, Aravind and O'Neill, Kate M. and Losert, Wolfgang and Papoian, Garegin A. and Giniger, Edward},
  editor = {Gupton, Stephanie},
  year = {2023},
  journal = {Mol. Biol. Cell},
  volume = {34},
  pages = {ar83},
  doi = {10.1091/mbc.e23-01-0003}
}

@article{forghani2023,
  title = {A New View of Axon Growth and Guidance Grounded in the Stochastic Dynamics of Actin Networks},
  author = {Forghani, Rameen and Chandrasekaran, Aravind and Papoian, Garegin and Giniger, Edward},
  year = {2023},
  journal = {Open Biol.},
  volume = {13},
  number = {6},
  pages = {220359},
  doi = {10.1098/rsob.220359}
}

@article{haas2006,
  title = {Chemokine {{Signaling Mediates Self-Organizing Tissue Migration}} in the {{Zebrafish Lateral Line}}},
  author = {Haas, Petra and Gilmour, Darren},
  year = {2006},
  journal = {Dev. Cell},
  volume = {10},
  number = {5},
  pages = {673--680},
  doi = {10.1016/j.devcel.2006.02.019}
}

@article{stoeckli2018,
  title = {Understanding Axon Guidance: Are We Nearly There Yet?},
  author = {Stoeckli, Esther T.},
  year = {2018},
  journal = {Development},
  volume = {145},
  number = {10},
  pages = {dev151415},
  doi = {10.1242/dev.151415}
}

@article{bornstein2021,
  title = {More than Movement: The Proprioceptive System as a New Regulator of Musculoskeletal Biology},
  author = {Bornstein, Bavat and Konstantin, Nitzan and Alessandro, Cristiano and Tresch, Matthew C and Zelzer, Elazar},
  year = {2021},
  journal = {Curr. Opin. Physiol.},
  volume = {20},
  pages = {77--89},
  doi = {10.1016/j.cophys.2021.01.004}
}

@article{blecher2017,
  title = {The {{Proprioceptive System Masterminds Spinal Alignment}}: {{Insight}} into the {{Mechanism}} of {{Scoliosis}}},
  shorttitle = {The {{Proprioceptive System Masterminds Spinal Alignment}}},
  author = {Blecher, Ronen and Krief, Sharon and Galili, Tal and Biton, Inbal E. and Stern, Tomer and Assaraf, Eran and Levanon, Ditsa and Appel, Elena and Anekstein, Yoram and Agar, Gabriel and Groner, Yoram and Zelzer, Elazar},
  year = {2017},
  journal = {Dev. Cell},
  volume = {42},
  number = {4},
  pages = {388--399.e3},
  doi = {10.1016/j.devcel.2017.07.022}
}

@article{hoffmann2026,
  title = {Postural control in an upright snake},
  author = {Hoffmann, Ludwig A. and Bryde, Petur and Davenport, Ian C. and Prasath, S Ganga and Jayne, Bruce C. and Mahadevan, L.},
  journal={J. R. Soc. Interface},
  year={2026},
  volume={23},
  number={235},
  pages = {20250314},
  doi = {10.1098/rsif.2025.0314}
}

@article{ladoux2017,
  title = {Mechanobiology of Collective Cell Behaviours},
  author = {Ladoux, Benoit and M{\`e}ge, Ren{\'e}-Marc},
  year = {2017},
  journal = {Nat. Rev. Mol. Cell Biol.},
  volume = {18},
  number = {12},
  pages = {743--757},
  doi = {10.1038/nrm.2017.98}
}

@article{naoki2019,
  title = {Noise-Resistant Developmental Reproducibility in Vertebrate Somite Formation},
  author = {Naoki, Honda and Akiyama, Ryutaro and Sari, Dini Wahyu Kartika and Ishii, Shin and Bessho, Yasumasa and Matsui, Takaaki},
  year = {2019},
  journal = {PLoS Comput. Biol.},
  volume = {15},
  number = {2},
  pages = {e1006579},
  doi = {10.1371/journal.pcbi.1006579}
}

@article{legoff2016,
  title = {Mechanical {{Forces}} and {{Growth}} in {{Animal Tissues}}},
  author = {LeGoff, Lo{\"i}c and Lecuit, Thomas},
  year = 2016,
  journal = {Cold Spring Harb. Perspect. Biol.},
  volume = {8},
  number = {3},
  pages = {a019232},
  doi = {10.1101/cshperspect.a019232}
}

@article{valet2022,
  title = {Mechanical Regulation of Early Vertebrate Embryogenesis},
  author = {Valet, Manon and Siggia, Eric D. and Brivanlou, Ali H.},
  year = 2022,
  journal = {Nat. Rev. Mol. Cell Biol.},
  volume = {23},
  number = {3},
  pages = {169--184},
  doi = {10.1038/s41580-021-00424-z}
}

@article{harmansa2021,
  title = {Forward and Feedback Control Mechanisms of Developmental Tissue Growth},
  author = {Harmansa, Stefan and Lecuit, Thomas},
  year = 2021,
  journal = {Cells Dev.},
  volume = {168},
  pages = {203750},
  doi = {10.1016/j.cdev.2021.203750}
}

@article{chen2021,
  title = {The {{Emerging Science}} of {{Interoception}}: {{Sensing}}, {{Integrating}}, {{Interpreting}}, and {{Regulating Signals}} within the {{Self}}},
  author = {Chen, Wen G. and Schloesser, Dana and Arensdorf, Angela M. and Simmons, Janine M. and Cui, Changhai and Valentino, Rita and Gnadt, James W. and Nielsen, Lisbeth and {Hillaire-Clarke}, Coryse St and Spruance, Victoria and Horowitz, Todd S. and Vallejo, Yolanda F. and Langevin, Helene M.},
  year = 2021,
  journal = {Trends Neurosci.},
  volume = {44},
  number = {1},
  pages = {3--16},
  doi = {10.1016/j.tins.2020.10.007}
}
\end{document}


\title{Supplementary Materials: How to grow a straight filament}
\author{L. A. Hoffmann}  
\affiliation{School of Engineering and Applied Sciences, Harvard University, Cambridge, Massachusetts 02138, USA}
\author{L. Mahadevan}
\email{lmahadev@g.harvard.edu}
\affiliation{School of Engineering and Applied Sciences, Department of Physics, and Department of Organismic and Evolutionary Biology, Harvard University, Cambridge, Massachusetts 02138, USA}
\date{\today}
\maketitle

\tableofcontents

\renewcommand{\theequation}{S\arabic{equation}}
\renewcommand{\thefigure}{S\arabic{figure}}
\renewcommand{\thesection}{S\Roman{section}}
\renewcommand{\thetable}{S\arabic{table}}

\newpage
\section{Parameters and Dynamic Equations}
\subsection{Table of Parameters}
\begin{table}[h]
\centering
\renewcommand{\arraystretch}{1.2}
\begin{tabular}{| c | c || c | c |}
\hline
Parameter/field & Meaning & Parameter/field & Meaning \\
\hline
$S$ & Elastic stretching coefficient & $w(s,t)$ & Out-of-line displacement \\
$B$ & Elastic bending coefficient & $\theta(s,t) = \partial_s w$ & Angle \\
$k$ & Coupling to substrate & $\theta_{\rm p}$ & Preferred angle \\
$\epsilon_{\rm e}(s,t)$ & Elastic strain & $\kappa(s,t) = \partial^2_s w$ & Curvature \\
$\kappa_{\rm e}(s,t)$ & Elastic curvature & $\chi_j(s,t)$ & Noise \\
$\epsilon_{\rm g}(s,t)$ & Strain growth & $f_i$ & Local feedback on strain growth \\
$\kappa_{\rm g}(s,t)$ & Curvature growth & $g_i$ & Local feedback on curvature growth \\
$u$ & In-line displacement & $\varphi_i$ & Nonlocal feedback on strain growth \\
$\epsilon = \partial_s u$ & Strain & $\gamma_i$ & Nonlocal feedback on curvature growth \\
\hline
\end{tabular}
\caption{\textbf{Parameters and fields.} We summarize the main parameters and fields, their meaning, as well as some relations.}
\label{tab:Parameters}
\end{table}
In Tab.~\ref{tab:Parameters}, we list the parameters and fields used in this work, together with their interpretation.

\subsection{Derivation of Equations}
To model the growing filament, we combine three key components: $(i)$ the filament is represented as an elongated quasi-one-dimensional elastic structure, with a specific expression for the free energy; $(ii)$ a growth law that dictates the precise manner in which growth occurs; and $(iii)$ a feedback law that allows internal and external signals to influence the growth over time and space. The remainder of this section provides a more detailed explanation of each of these components. We assume that the filament behaves as a purely elastic structure. In the absence of growth, there is a unique, time-independent ground state, and perturbations to this ground state incur an energetic cost. We consider the linearized free energy
\begin{equation}
\mathcal{F} = \int {\rm d}A\; \frac{S}{2} \epsilon_{\rm e}(s,t)^2 + \frac{B}{2} \kappa_{\rm e}(s,t)^2 + \frac{k}{2} w(s,t)^2 \;,
\end{equation}
which is the sum of the elastic stretching energy, the elastic bending energy, and a Hookean coupling to a substrate, with elastic constants $S$, $B$, and $k$, respectively. The integration is over the length $L(t)$ of the filament as well as the width, which is assumed to be much smaller than the length such that the structure is quasi-one-dimensional. In the absence of growth, $\epsilon_{\rm e} = \partial_s u$ and $\kappa_{\rm e} = \partial_s^2 w$ are the equilibrium strain and curvature, with $u(s,t)$ the in-line and $w(s,t)$ the out-of-line deformation. Through the third term, the filament is coupled to a substrate taken to be at $w = 0$ and therefore breaking translation invariance.

The filament can undergo either in-line or out-of-line growth. For a quasi-one-dimensional structure with a finite but small width, in-line growth occurs when both sides of the filament grow at the same rate, leading to an increase in the length of the flat structure. In contrast, out-of-line growth occurs when one side grows faster than the other, resulting in unequal lengths on the two sides and causing the filament to bend. In both cases, we model these two modes of growth as modifications to the ground state of the filament. Namely, denoting by $\epsilon_\text{g}$ the strain due to in-line growth and by $\kappa_\text{g}$ the curvature due to out-of-line growth, we can write~\cite{efrati2009,liang2009,al-mosleh2023}
\begin{subequations}
\label{eq:GrowStCurv}
\begin{align}
\epsilon(s,t) &= \epsilon_{\rm e}(s,t) + \epsilon_{\rm g}(s,t) = \partial_s u(s,t) \;, \\
\kappa(s,t) &= \kappa_{\rm e}(s,t) + \kappa_{\rm g}(s,t) = \partial_s^2 w(s,t) \;.
\end{align}
\end{subequations}
Comparing these expressions with those above shows how growth modifies the ground state. Therefore, depending on the precise form of the growth $\{\epsilon_{\rm g}, \kappa_{\rm g}\}$ the elastic energy can either increase or decrease over time. Note that if growth does not induce any change in strain or curvature, it does not contribute to the free energy and, as a result, does not appear in this framework. For instance, one could imagine an elastic structure growing homogeneously and isotropically in-line, where no growth-induced strain or curvature is generated. In such a case, the elastic energy of the filament remains unaffected, and growth does not explicitly enter the equations presented here.
Minimizing the free energy with respect to $u$ and $w$ yields two equations:
\begin{subequations}
\begin{align}
&\frac{\delta \mathcal{F}}{\delta u} =  -S \partial_s \left(\partial_s u(s,t) - \epsilon_{\rm g}(s,t)\right) = 0 \;, \\
&\frac{\delta \mathcal{F}}{\delta w} =  B \partial^2_s \left(\partial_s^2 w(s,t) - \kappa_{\rm g}(s,t)\right) + k w = 0 \;.
\end{align}
\label{eq:EOMSI}
\end{subequations}
It thus remains to specify the growth field. In principle, one could impose an arbitrary growth field in both space and time. However, such growth could lead to a state where internal stresses accumulate over time due to incompatible growth, meaning that different regions of the filament grow in an uncoordinated manner. To prevent this, it is expected that the growth fields at a given time would be influenced by the stresses present in the system at that time (or possibly also earlier times). In this way, growth could be reduced, for example, in regions of high stress, allowing for self-regulation of the growth and stress fields, thereby preventing the accumulation of internal stresses. The feedback laws we consider can be written in their most general form as:
\begin{subequations}
\begin{align}
\partial_t \epsilon_{\rm g}(s,t) = &\int \text{d}s' \text{d}t' \left[F_{\epsilon} \epsilon(s',t') + F_{\theta} \sin\left(\theta(s',t')-\theta_\text{p}\right)  + F_{\kappa} \kappa(s',t') \right] + \chi_\epsilon(s,t) \;, \\
\partial_t \kappa_{\rm g}(s,t) = &\int \text{d}s' \text{d}t' \left[G_{\epsilon} \epsilon(s',t') + G_{\theta} \sin\left(\theta(s',t')-\theta_\text{p}\right)  + G_{\kappa} \kappa(s',t') \right] + \chi_\kappa(s,t) \;,
\end{align}
\label{eq:GeneralFeedbackSI}
\end{subequations}
where $F_i \equiv F_i(t,t',s,s')$ and $G_i \equiv G_i(t,t',s,s')$ are integral kernels allowing for the growth fields at $(s,t)$ to be affected by fields at different positions $s'$ (nonlocal feedback) and at different times $t'$ (time-delayed feedback). Furthermore, the terms proportional to $F_\theta$ and $G_\theta$ specify that there is a preferred orientation of the filament. $\theta_\text{p}$ is a free parameter that determines the angle of the preferred orientation while $\theta(s',t')$ is the angle of the filament at point $(s',t')$.

In the absence of an elastic substrate ($k = 0$), combining Eqs.~\eqref{eq:EOMSI} and \eqref{eq:GeneralFeedbackSI} yields
\begin{subequations}
\label{eq:CombinedFullEq}
\begin{align}
\partial_t \partial_s u(s,t) &= \int \text{d}s' \text{d}t' \left[F_{\epsilon} \epsilon(s',t') + F_{\theta} \sin\left(\theta(s',t')-\theta_\text{p}\right)  + F_{\kappa} \kappa(s',t') \right] + \chi_\epsilon(s,t) \;,  \\
\partial_t \partial^2_s w(s,t) &= \int \text{d}s' \text{d}t' \left[G_{\epsilon} \epsilon(s',t') + G_{\theta} \sin\left(\theta(s',t')-\theta_\text{p}\right)  + G_{\kappa} \kappa(s',t') \right] + \chi_\kappa(s,t) \;.
\end{align}
\end{subequations}

In the simplest case of local and instantaneous feedback, $F_i(t,t',s,s') = - f_i \delta(s - s') \delta(t-t')$ and $G_i(t,t',s,s') = - g_i \delta(s - s') \delta(t-t')$, with parameters $f_i$ and $g_i$, such that the growth at $(s,t)$ is affected only by the conditions at the same time and point in space.
Then, Eqs.~\eqref{eq:CombinedFullEq} reduce to
\begin{subequations}
\label{eq:LocalEquations}
\begin{align}
\partial_t \partial_s u(s,t) \simeq &- f_{\epsilon} \epsilon(s,t) - f_{\theta} \sin(\theta(s,t) - \theta_\text{p}) - f_{\kappa} \kappa(s,t) + \chi_\epsilon(s,t) \;, \\
\partial_t \partial_s^2 w(s,t) \simeq &- g_{\epsilon} \epsilon(s,t) - g_{\theta} \sin(\theta(s,t) - \theta_\text{p}) - g_{\kappa} \kappa(s,t) + \chi_\kappa(s,t) \;.
\end{align}
\end{subequations}

\section{A Paradigmatic Example: Growing Plant Shoots}
In recent years, plants have served as a paradigmatic example where the interplay of growth, feedback, and noise has been investigated in great detail. We describe here in more detail the feedback mechanisms that have been found to be relevant during plant morphogenesis to regulate development. Furthermore, we show how our model connects to the classic work of Bastien et al.~\cite{bastien2013}, where proprioceptive feedback and orientation sensing (graviotropism, sensing of the gravitational field) were modeled to explain the experimentally observed growth dynamics of plants.

\subsection{Details on Plant Feedback System}
The growth of animal tissues primarily occurs through cell division, but in plant tissues, growth can also be driven by the expansion of cell size~\cite{hong2018}. Recent studies tracking cell shape and growth rates have provided insights into the stochastic nature of cellular growth~\cite{moulia2021}. For instance, it has been observed that neighboring cells can exhibit significantly different expansion rates~\cite{hong2018,armon2021}, leading to a highly noisy growth field. In fact, it has been proposed that noise may be essential for robust development, with abnormalities arising in the absence of such noisy growth~\cite{moulia2021}. As noted, this raises the question of how shape dynamics are controlled at the macroscopic scale. In recent years, sensing and feedback mechanisms have emerged as promising candidates for regulating these processes~\cite{shraiman2005, hong2018, damavandi2019, moulia2021}. 

Sensing involves the processing of external and internal information. It has been demonstrated, for example, that amyloplasts are involved in sensing the gravitational field~\cite{okamoto2015, berut2018}, enabling the plant to perceive ``up'' and ``down'' even in the absence of sunlight, which is also utilized as a source of information (phototropism). Other external forces, such as wind, can also influence growth.
Internal sensing refers to the ``measurement'' of internal strains and curvature. Cortical microtubules (CMTs) within plant cells have been shown to rearrange their orientation in response to the forces acting on a cell. Specifically, when a cell is stretched or compressed anisotropically, the CMTs align with the direction of maximal tension, thereby increasing the stiffness of the cell in that direction. In this way, internal stresses influence cellular growth~\cite{hamant2019}. Furthermore, actin cables oriented along a growing stem have been implicated in sensing curvature, allowing the growing organism to infer information about its own shape, a process known as proprioception~\cite{bastien2013, okamoto2015, chelakkot2017, meroz2019, moulia2021}. 
These sensing mechanisms are connected to feedback systems that modulate growth accordingly. For instance, the realignment of CMTs described above is thought to regulate growth rate and direction in response to local tissue stresses. Additionally, the plant hormone auxin has been identified as a key regulator of growth, with higher (or lower) concentrations promoting (or inhibiting) growth~\cite{nakayama2012, jonsson2022}. Auxin has been shown to be released in response to shape deformations, thereby self-correcting deviations from the desired growth pattern and, for example, maintaining sepal flatness~\cite{xu2024}. In fact, active transport of auxin through the tissues, mediated by the cell-membrane-localized PIN1 protein~\cite{hong2018}, facilitates growth regulation at the supercellular level. In this manner, sensing and feedback mechanisms are thought to ensure robust development, although many details of these processes remain unknown.

\subsection{Bending of a Growing Plant Shoot}
\begin{figure*}
\centering
\includegraphics[width=\textwidth]{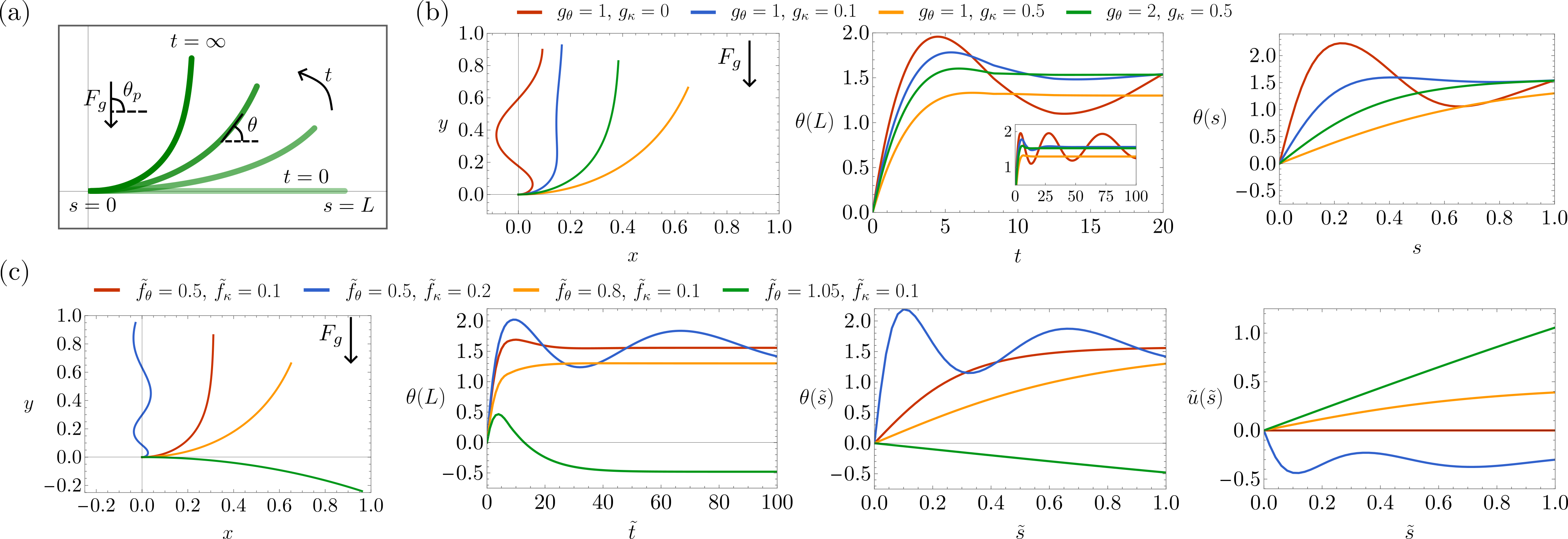} 
\caption{\textbf{Reorientation toward the gravitational field.} (a) Sketch of the initial horizontal condition and the time evolution as the filament bends upward. (b) For different values of the parameters $g_{\theta}$ and $g_{\kappa}$ we show the shape in real space (at $t = 20$), the time evolution of the angle at the tip $\theta(L)$, and the angle along the filament $\theta(s)$ at $t = 20$ as found when solving Eq.~\eqref{eq:BastiensModel}. The inset in the second panel highlights that for $g_{\kappa} = 0$ the filament oscillates indefinitely.  
(c) For different values of the parameters $\tilde{f}_{\theta}$ and $\tilde{f}_{\kappa}$ (as defined in the text) and for $\tilde{g}_{\kappa} = 0.2$ we show the same information as in (b) as well as the strain along the filament.}
\label{fig:Bastien}
\end{figure*}
Bastien et al.~\cite{bastien2013} studied (in our notation) the following equation:
\begin{equation}
\label{eq:BastiensModel}
\partial_t \kappa(s,t) = -  g_{\theta} \sin(\theta(s,t) - \theta_\text{p}) - g_{\kappa} \kappa(s,t) \;.
\end{equation}
It is thus the deterministic limit of Eqs.~\eqref{eq:LocalEquations} where $f_i = 0$ and $g_\epsilon = 0$. This model has been used to explain the observed upward bending of an initially horizontal plant shoot towards the vertical direction, induced by the presence of the gravitational field, see Fig.~\ref{fig:Bastien}(a). The shoot is fixed at the origin such that its orientation there cannot change, thus $\theta(s = 0, t) = 0$. We show the solution of Eq.~\eqref{eq:BastiensModel} for selected parameter values in Fig.~\ref{fig:Bastien}(b). It is found that for a certain range of parameter values, the plant shoot reaches a steady state where it is bent upwards, with the precise shape at long times determined by the choice of these parameters. If $g_{\theta} = 0$, the plant remains horizontal and does not bend. However, if $g_{\kappa} = 0$ but $g_{\theta} \neq 0$, the filament bends but does not reach a steady state. Instead, it continuously oscillates with decreasing amplitude. To explain this behavior, we analyze the stability of the straight state. Namely, linearizing the above equation about $\theta = \theta_\text{p} + \delta \theta = \pi/2 + \delta \theta$ yields
\begin{equation}
\partial_t \partial_s \delta \theta(s,t)= -  g_{\theta} \delta \theta(s,t) - g_{\kappa} \partial_s \delta \theta(s,t) \;,
\end{equation}
where $\kappa = \partial_s \theta$ was used. To investigate for which parameters the system reaches a steady state at long times we substitute
\begin{equation}
\label{eq:LinInstTheta}
\delta \theta(s,t) = \theta_0 e^{i q s + \lambda t}
\end{equation}
to find
\begin{equation}
\lambda = -g_{\kappa} + i \frac{g_{\theta}}{q} \;.
\end{equation}
If the real part of $\lambda$ is positive, the system is unstable as the expression in Eq.~\eqref{eq:LinInstTheta} diverges exponentially for large $t$ such that $\delta \theta$ grows without bound. On the other hand, if it is negative, the system is linearly stable; a small perturbation of the initial state will decay in time and the system will relax to its ground state for large times as $\delta \theta \to 0$. If the eigenvalues have a nonzero imaginary part, the system is oscillatory.
Therefore, if $g_{\kappa} > 0$, $\text{Re}\left[\lambda\right] = -g_{\kappa} < 0$ and the system is stable. On the other hand, if $g_{\kappa} = 0$, then $\lambda$ is purely imaginary and the system is oscillatory. This explains the observation in Fig.~\ref{fig:Bastien}(b) that the filament continues to oscillate indefinitely. The proprioceptive term $g_{\kappa} \kappa$ is required for reaching a stable state in finite time.

We can now study the same scenario in our more general setting where we include the dynamics of the strain field, cf. Eqs.~\eqref{eq:LocalEquations}.
We reduce the number of parameters by defining dimensionless coordinates for time $\tilde{t} = f_{\epsilon} t$ and space $\tilde{s} = s g_{\theta}/f_{\epsilon}$. Furthermore, we can then define the dimensionless parameters $\tilde{g}_{\kappa} = g_{\kappa}/f_{\epsilon}$, $\tilde{g}_{\epsilon} = g_{\epsilon}/g_{\theta}$, $\tilde{f}_{\theta} = f_{\theta} \tilde{g}_{\epsilon}/f_{\epsilon}$, and $\tilde{f}_{\kappa} = f_{\kappa} \tilde{g}_{\epsilon} g_{\theta}/f_{\epsilon}^2$ such that there are only three free parameters -- namely $\tilde{f}_{\theta}$, $\tilde{f}_{\kappa}$, and $\tilde{g}_{\kappa}$ -- and the deterministic limit of Eqs.~\eqref{eq:LocalEquations} reads
\begin{subequations}
\label{eq:LocalNonDimEqs}
\begin{align}
\partial_{\tilde{t}} \partial_{\tilde{s}} \tilde{u}(\tilde{s},\tilde{t}) &\simeq - \partial_{\tilde{s}} \tilde{u}(\tilde{s},\tilde{t}) - \tilde{f}_{\theta} \sin\left[\theta(\tilde{s},\tilde{t}) - \theta_p \right]  - \tilde{f}_{\kappa} \partial_{\tilde{s}} \theta(\tilde{s},\tilde{t}) \;, \\
\partial_{\tilde{t}} \partial_{\tilde{s}} \theta(\tilde{s},\tilde{t}) &\simeq - \partial_{\tilde{s}} \tilde{u}(\tilde{s},\tilde{t}) - \sin\left[\theta(\tilde{s},\tilde{t}) - \theta_p \right]  - \tilde{g}_{\kappa} \partial_{\tilde{s}} \theta(\tilde{s},\tilde{t})  \;,
\end{align}
\end{subequations}
where $\tilde{u} = u g_{\epsilon}/f_{\epsilon}$.
We impose $\theta(\tilde{s}, \tilde{t} = 0) = 0$, $\theta(\tilde{s} = 0, \tilde{t}) = 0$ as the boundary conditions as before and furthermore set $\tilde{u}(\tilde{s} = 0, \tilde{t}) = 0$, and $\tilde{u}(\tilde{s}, \tilde{t} = 0) = 0$. Results for selected parameter values are shown in Fig.~\ref{fig:Bastien}(c). As in the previous subsection we find that the system reaches a steady state in finite time only for some regions of parameter space. However, this region is different from before and we find, for example, an oscillating state even if $g_{\kappa} \neq 0$. Furthermore, for some parameter values, the filament bends \textit{downwards}.

\section{Local, instantaneous feedback}
\label{sec:LocalFeedback}
We now return to the general setting and ask: starting from an initially straight filament with $\theta(s,t = 0) = \theta_\text{p}$, under what conditions does the filament remain straight? For small deviations from a reference state, we approximate to first order $\epsilon \simeq \partial_s u$, $\delta \theta \simeq \partial_s w$, $\kappa \simeq \partial_s^2 w$, and $\sin \delta \theta \simeq \delta \theta$, so that Eqs.~\eqref{eq:CombinedFullEq} become
\begin{subequations}
\begin{align}
\partial_t \partial_s u(s,t) &= \int \text{d}s' \text{d}t' \left[F_{\epsilon} \partial_{s'} u(s',t') + F_{\theta} \partial_{s'} w(s',t')  + F_{\kappa} \partial^2_{s'} w(s',t') \right] + \chi_\epsilon(s,t) \;,  \\
\partial_t \partial^2_s w(s,t) &= \int \text{d}s' \text{d}t' \left[G_{\epsilon} \partial_{s'} u(s',t') + G_{\theta} \partial_{s'} w(s',t')  + G_{\kappa} \partial^2_{s'} w(s',t') \right] + \chi_\kappa(s,t) \;.
\end{align}
\end{subequations}
It is now transparent that the feedback laws we chose are the lowest-order terms in the derivatives of $(u,w)$, with the zero-derivative terms vanishing due to translation symmetry. This can be written more compactly as
\begin{equation}
\partial_t \partial_s \bm{V}(s,t) = \int \bm{G}(s, s', t, t') \cdot \bm{V}(s',t') \, \text{d}s' \text{d}t' + \bm{\chi} \;,
\label{eq:GeneralVecEq}
\end{equation}
where $\bm{V}(s,t) = (u(s,t), \theta(s,t))^\top$, $\bm{\chi}(s,t) = (\chi_\epsilon(s,t), \chi_\kappa(s,t))^\top$, and
\begin{equation}
\bm{G}(s, s', t, t') = \begin{pmatrix} F_{\epsilon} \partial_{s'} & F_{\theta} + F_{\kappa} \partial_{s'} \\ G_{\epsilon} \partial_{s'} & G_{\theta} + G_{\kappa} \partial_{s'}  \end{pmatrix} \;.
\end{equation}
For local feedback, this reduces to:
\begin{equation}
\bm{G} = - \begin{pmatrix} f_{\epsilon} \partial_s & f_{\theta} + f_{\kappa}  \partial_s   \\ g_{\epsilon} \partial_s & g_{\theta} + g_{\kappa}  \partial_s   \end{pmatrix} \;,
\end{equation}
and we analyze this case now. First, we consider the deterministic equations and afterwards investigate how the presence of noise modifies the results. 

\subsection{Linear stability of deterministic equations}
We first analyze the linear stability of the dimensional equations and then nondimensionalize them.

\subsubsection{Eigenvalue formulas}
Using the ansatz
\begin{equation}
u = u_0 e^{i q s + \lambda t} \qquad \text{and} \qquad \theta = \theta_0 e^{i q s + \lambda t}
\end{equation}
in
\begin{subequations}
\label{eq:LocalDimEqs}
\begin{align}
\partial_{t} \partial_{s} u(s,t) &\simeq - f_{\epsilon} \partial_{s} u(s,t) - f_{\theta} \theta(s,t)  - f_{\kappa} \partial_{s} \theta(s,t) \;, \\
\partial_{t} \partial_{s} \theta(s,t) &\simeq - g_{\epsilon} \partial_{s} u(s,t) - g_{\theta} \theta(s,t)  - g_{\kappa} \partial_{s} \theta(s,t)  \;,
\end{align}
\end{subequations}
cf. Eqs.~\eqref{eq:GeneralVecEq}, it is possible to write this system of equations compactly as
\begin{subequations}
\begin{equation}
\left[\lambda \mathbf{1} - \bm{K} \right] \cdot \bm{V} = 0 \;,
\end{equation}
where
\begin{equation}
\label{eq:KExpression}
\bm{K}(q) = 
    \begin{pmatrix} 
    - f_{\epsilon} & -f_{\kappa}+i f_{\theta}/q    \\ 
    - g_{\epsilon} & - g_{\kappa} + i g_{\theta}/q 
    \end{pmatrix} \;.
\end{equation}
\end{subequations}
To determine the region of (in)stability we solve the equation $\det\left[\lambda \mathbf{1} - \bm{K} \right] = 0$ for $\lambda$ and it directly follows that
\begin{align}
\label{eq:LambdaExpress}
\lambda_{1,2}(q) &= \frac{1}{2} \left(\text{tr}[\bm{K}] \pm \sqrt{\text{tr}[\bm{K}]^2 - 4\, \det[\bm{K}]} \right) \nonumber \\
& = - \frac{f_{\epsilon} + g_{\kappa}}{2} + \frac{i g_{\theta}}{2q} \pm \frac{i \sqrt{g_{\theta}^2 + 2i[2 f_{\theta} g_{\epsilon} + g_{\theta} (g_{\kappa} - f_{\epsilon})]q - [4 f_{\kappa} g_{\epsilon} + (g_{\kappa} - f_{\epsilon})^2]q^2}}{2q} \;.
\end{align}
We show some examples below, after rescaling the expressions to reduce the number of parameters. Note that $g_{\epsilon}$ only appears in the combination $f_{\theta} g_{\epsilon}$ and $f_{\kappa} g_{\epsilon}$ and thus we can define $\bar{f}_\theta = f_{\theta} g_{\epsilon}/g_{\theta}$ and $\bar{f}_\kappa = f_{\kappa} g_{\epsilon}$, where we have included $g_{\theta}$ in the definition of $\bar{f}_{\theta}$ for future convenience, as will become evident below.
\begin{figure*}
\centering
\includegraphics[width=1\textwidth]{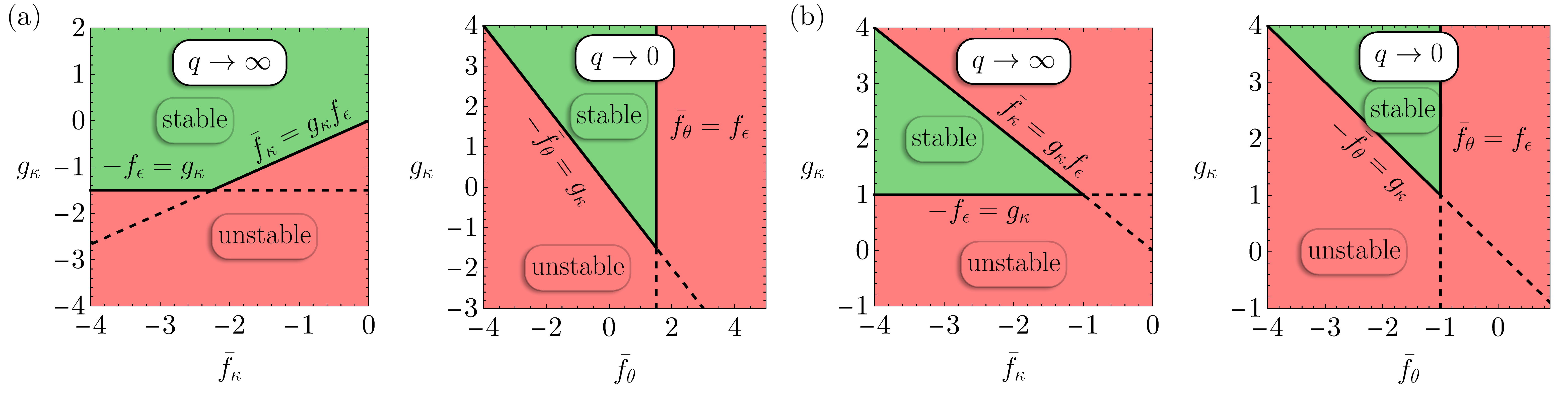}
\caption{\textbf{Stability diagrams for local feedback.} Two examples of stability diagrams for the inequalities in Eqs.~\eqref{eq:InequalDim} found from the asymptotic expansion for $q \to 0, \infty$. The only free parameter here is $f_{\epsilon}$ and we set (a) $f_{\epsilon} = 1.5$ and (b) $f_{\epsilon} = -1$. In all cases the region of stability is bounded by two straight lines. Note that the sign of $f_{\epsilon}$ determines the sign of the slope of the line $\bar{f}_\kappa = g_{\kappa} f_{\epsilon}$. Here, $\bar{f}_\theta = f_{\theta} g_{\epsilon}/g_{\theta}$ and $\bar{f}_\kappa = f_{\kappa} g_{\epsilon}$.}
\label{fig:StabDia}
\end{figure*}

\subsubsection{Asymptotic limits of eigenvalues}
We now consider the asymptotic limits $q \to 0,\,\infty$ to analyze the stability of the system at long and short length scales.
For short length scales, expanding about $q = \infty$ gives the following lowest-order expressions for the real and imaginary parts: 
\begin{subequations}
\label{eq:EVLocalInfty}
\begin{align}
\lambda_{1,2}^{\text{RS}} &= \text{Re}\left[\lambda_{1,2}\right]_{q \to \infty} = -\frac{f_{\epsilon} + g_{\kappa}}{2} \pm \frac{1}{2} \text{Re} \left[ \sqrt{4 \bar{f}_\kappa + \left(g_{\kappa} - f_{\epsilon}\right)^2} \right]+ \mathcal{O}\left(q^{-1}\right) \;, \\
\lambda_{1,2}^{\text{IS}} &= \text{Im}\left[\lambda_{1,2}\right]_{q \to \infty} = \mp \frac{1}{2} \text{Im} \left[\sqrt{4 \bar{f}_\kappa +(g_{\kappa} - f_{\epsilon})^2}\right] + \mathcal{O}\left(q^{-1}\right) \;.
\end{align}
\end{subequations}
If the real part of the eigenvalues is positive, the system is unstable. On the other hand, if it is negative, the system is linearly stable; a small perturbation of the initial state will decay and the system will relax back to it. If the eigenvalues have a nonzero imaginary part, the system is oscillatory. We can thus analyze the respective expressions to derive conditions for the parameters for the linear (in)stability of the system. From $\lambda_{1,2}^{\text{RS}} \le 0$ we find the condition
\begin{equation}
-f_{\epsilon} - g_{\kappa} \le \text{Re} \left[ \sqrt{4 \bar{f}_\kappa + \left(g_{\kappa} - f_{\epsilon}\right)^2} \right] \le f_{\epsilon} + g_{\kappa}  \;.
\end{equation}
We thus find the conditions $-f_{\epsilon} \le  g_{\kappa}$ and $\bar{f}_\kappa \le g_{\kappa} f_{\epsilon}$ for stability. Note that this reduces to the condition $g_{\kappa} \ge 0$ in the model of Bastien et al. described in the previous section. For the imaginary part we find that $\lambda_{1,2}^{\text{IS}} \to 0$ if $4 \bar{f}_\kappa + (g_{\kappa} - f_{\epsilon})^2 \ge 0$.

\begin{figure*}
\centering
\includegraphics[width=1.\textwidth]{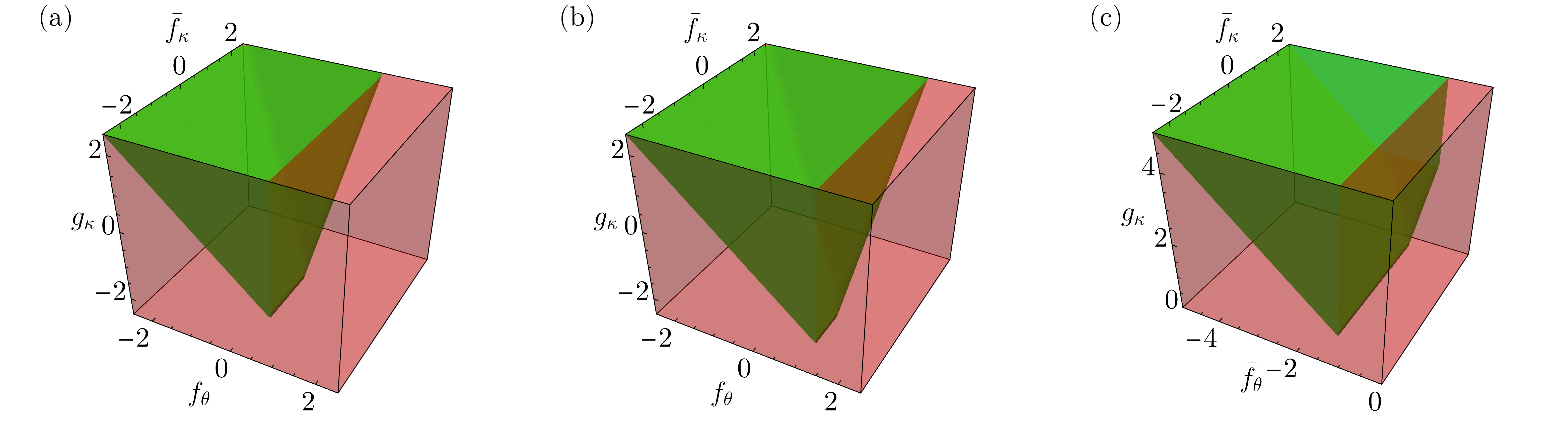}
\caption{\textbf{3D stability diagram for local feedback.} Three examples of three-dimensional stability diagrams for the inequalities in Eq.~\eqref{eq:InequalDim}. Despite these inequalities being found from considering $q \to 0,\infty$, due to the monotonicity of $\text{Re } \lambda(q)$ the stable region in the stability diagrams here is valid for all $q$. For the only free parameter $f_{\epsilon}$ we choose (a) $f_{\epsilon} = 1$, (b) $f_{\epsilon} = 1.5$, and (c) $f_{\epsilon} = -1$. The diagrams of Fig.~3 in the main text and of Fig.~\ref{fig:StabDia} are thus two-dimensional slices of these three-dimensional diagrams.}
\label{fig:StabDia3D}
\end{figure*}

On the other hand, for large lengths we find by expanding about $q = 0$:
\begin{subequations}
\begin{align}
\lambda_{1,2}^{\text{RL}} &= \text{Re}\left[\lambda_{1,2}\right]_{q \to 0} = -  \frac{g_{\kappa} + f_{\epsilon}}{2} \pm \frac{(g_{\kappa} - f_{\epsilon}) + 2\bar{f}_\theta}{2} + \mathcal{O}\left(q^2\right) \\
\lambda_{1,2}^{\text{IL}} &= \text{Im}\left[\lambda_{1,2}\right]_{q \to 0} = \frac{g_{\theta} \mp g_{\theta}}{2q} + \mathcal{O}\left(q\right)\;,
\end{align}
\end{subequations}
Hence, in the limit $q \to 0$, one imaginary part vanishes while the other diverges, independently of the parameter values. Furthermore, from $\lambda_{1}^{\text{RL}} \le 0$ we obtain the condition $\bar{f}_\theta \le f_{\epsilon}$ while from
$\lambda_{2}^{\text{RL}} \le 0$ we find $-\bar{f}_\theta \le g_{\kappa}$. We summarize the conditions for the real parts of the eigenvalues to be nonpositive:
\begin{subequations}
\label{eq:InequalDim}
\begin{align}
&q \to 0: \qquad -\frac{f_{\theta} g_{\epsilon}}{g_{\theta}} \equiv -\bar{f}_\theta \le g_{\kappa} \qquad \text{and} \qquad \frac{f_{\theta} g_{\epsilon}}{g_{\theta}} \equiv \bar{f}_\theta \le f_{\epsilon} \;, \\
&q \to \infty: \qquad -f_{\epsilon} \le g_{\kappa} \qquad \text{and} \qquad f_{\kappa} g_{\epsilon} \equiv \bar{f}_\kappa \le g_{\kappa} f_{\epsilon} \;.
\end{align}
\end{subequations}
Examples of stability diagrams obtained from these inequalities are shown in Fig.~\ref{fig:StabDia}. Note that in Fig.~\ref{fig:StabDia} at the ``corner'', where the two lines intersect, the real part of the eigenvalue vanishes identically.
Furthermore, due to the monotonicity of $ \text{Re }\lambda(q)$, the inequalities found from asymptotic considerations here actually determine the stability for all $q$. We thus have three-dimensional stability diagrams, see Fig.~\ref{fig:StabDia3D} for some examples.

\subsection{Non-dimensional expressions}
We now convert the above expressions into the dimensionless form.

\subsubsection{Eigenvalue formulas}
Using the rescaled parameters defined above, $\tilde{g}_{\kappa} = g_{\kappa}/f_{\epsilon}$, $\tilde{g}_{\epsilon} = g_{\epsilon}/g_{\theta}$, $\tilde{f}_{\theta} = f_{\theta} \tilde{g}_{\epsilon}/f_{\epsilon}$, and $\tilde{f}_{\kappa} = f_{\kappa} \tilde{g}_{\epsilon} g_{\theta}/f_{\epsilon}^2$, as well as the dimensionless variables $\tilde{q} = q f_{\epsilon}/g_{\theta}$ and $\tilde{\lambda} = \lambda/f_{\epsilon}$, we can write $\left[\tilde{\lambda} \mathbf{1} - \tilde{\bm{K}} \right] \cdot \tilde{\bm{V}} = 0$ with
\begin{equation}
\label{eq:KExpressionNonDim}
\tilde{\bm{K}}(\tilde{q}) = 
- \begin{pmatrix} 1 & \frac{\tilde{f}_{\kappa} - i \tilde{f}_{\theta}/\tilde{q}}{\tilde{g}_{\epsilon}} \\ \tilde{g}_{\epsilon} & \tilde{g}_{\kappa} - \frac{i}{\tilde{q}} \end{pmatrix}
\end{equation}
such that the eigenvalues are given by
\begin{equation}
\label{eq:LambdaExpressNonDim}
\tilde{\lambda}_{1,2} = - \frac{1 + \tilde{g}_{\kappa}}{2} + \frac{i}{2 \tilde{q}} \pm \frac{i \sqrt{1 + 2i(\tilde{g}_{\kappa}+2 \tilde{f}_{\theta} - 1)\tilde{q} - [4 \tilde{f}_{\kappa} + (\tilde{g}_{\kappa} - 1)^2]\tilde{q}^2}}{2\tilde{q}} \;.
\end{equation}
\begin{figure*}
\centering
\includegraphics[width=\textwidth]{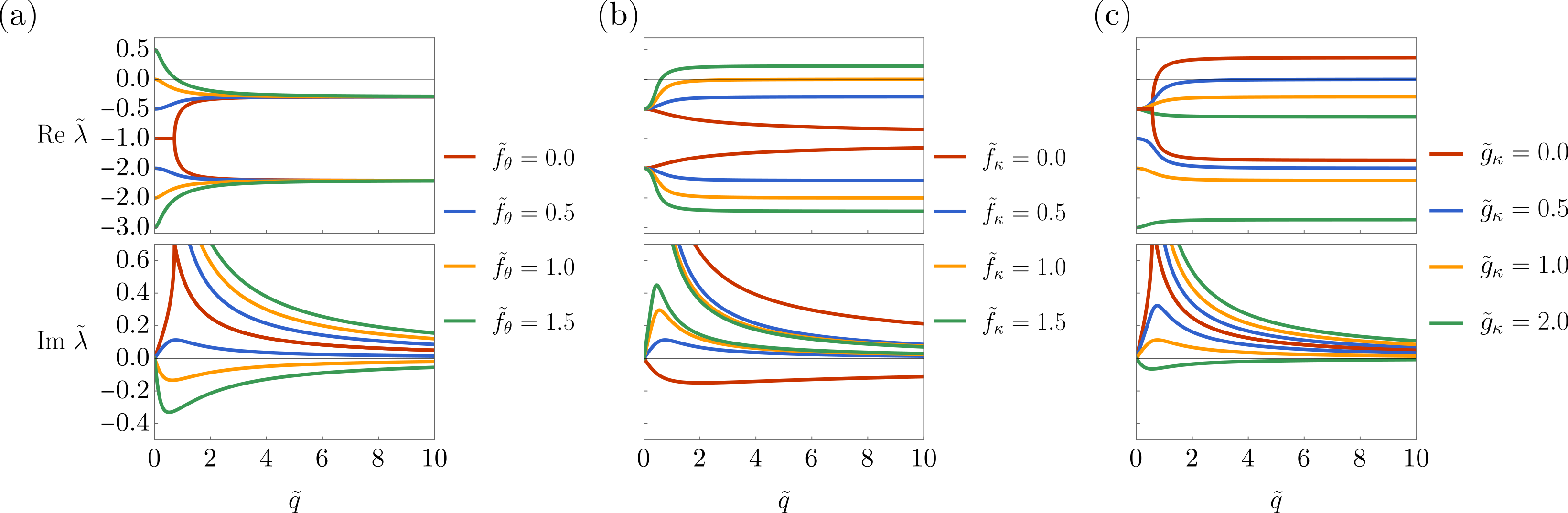}
\caption{\textbf{Eigenvalues for local feedback.} We show the real (top row) and imaginary (bottom row) parts of the eigenvalues in Eq.~\eqref{eq:LambdaExpressNonDim} for different parameter values. We vary $\tilde{f}_{\theta}$ in (a), $\tilde{f}_{\kappa}$ in (b), and $\tilde{g}_{\kappa}$ in (c), and set $\tilde{f}_{\theta} = 0.5$, $\tilde{f}_{\kappa} = 0.5$, and $\tilde{g}_{\kappa} = 1$ when the respective parameters are not varied.
}
\label{fig:Eigenvalues}
\end{figure*}
Some examples are shown in Fig.~\ref{fig:Eigenvalues}. 

\subsubsection{Asymptotic limits of eigenvalues}
The asymptotic limits are found to be
\begin{subequations}
\label{eq:EVLocalLimits}
\begin{align}
\tilde{\lambda}_{1,2}^{\text{RS}} &= \text{Re}\left[\tilde{\lambda}_{1,2}\right]_{\tilde{q} \to \infty} = -\frac{1 + \tilde{g}_{\kappa}}{2}\pm \frac{1}{2} \text{Re} \left[ \sqrt{4 \tilde{f}_{\kappa} + \left(\tilde{g}_{\kappa} - 1\right)^2} \right]+ \mathcal{O}\left(\tilde{q}^{-1}\right) \;, \\
\tilde{\lambda}_{1,2}^{\text{IS}} &= \text{Im}\left[\tilde{\lambda}_{1,2}\right]_{\tilde{q} \to \infty} = \mp \frac{1}{2} \text{Im} \left[\sqrt{4 \tilde{f}_{\kappa} + (\tilde{g}_{\kappa} - 1)^2}\right] + \mathcal{O}\left(\tilde{q}^{-1}\right) \;, \\
\tilde{\lambda}_{1,2}^{\text{RL}} &= \text{Re}\left[\tilde{\lambda}_{1,2}\right]_{\tilde{q} \to 0} = -  \frac{\tilde{g}_{\kappa} + 1}{2} \mp \frac{1 - 2\tilde{f}_{\theta} - \tilde{g}_{\kappa}}{2} + \mathcal{O}\left(\tilde{q}^2\right) \;, \\
\tilde{\lambda}_{1,2}^{\text{IL}} &= \text{Im}\left[\tilde{\lambda}_{1,2}\right]_{\tilde{q} \to 0} = \frac{1 \mp 1}{2\tilde{q}} + \mathcal{O}\left(\tilde{q}\right)\;,
\end{align}
\end{subequations}
Now, note that $\text{Re } \tilde{\lambda} < 0$ corresponds to $\text{Re } \lambda < 0$ only if $f_{\epsilon} > 0$. On the other hand, if $f_{\epsilon} < 0$, then we have to consider $\text{Re } \tilde{\lambda} > 0$ to find the parameter space where $\text{Re } \lambda < 0$. Thus, while the number of parameters is reduced to three, the linear instability analysis is slightly less convenient. In the case $f_{\epsilon} > 0$ we thus find the following inequalities: $\tilde{\lambda}_{1}^{\text{RS}} \le 0$ if $\tilde{g}_{\kappa} \ge -1$ and $\tilde{f}_{\kappa} \le \tilde{g}_{\kappa}$. If these (in)equalities are fulfilled, we always have $\tilde{\lambda}_{2}^{\text{RS}} \le 0$ as well and there are thus no additional constraints. For the imaginary part, we find that $\tilde{\lambda}_{1,2}^{\text{IS}} \to 0$ if $4 \tilde{f}_{\kappa} + (\tilde{g}_{\kappa} - 1)^2 > 0$. On the other hand, $\tilde{\lambda}_{1,2}^{\text{IL}} \to 0, \infty$ always, independent of parameters. Furthermore, from $\tilde{\lambda}_{1}^{\text{RL}} \le 0$ we obtain the condition $1 \ge \tilde{f}_{\theta}$ while from
$\tilde{\lambda}_{2}^{\text{RL}} \le 0$ we find $\tilde{g}_{\kappa} + \tilde{f}_{\theta} \ge 0$.
To summarize:
\begin{subequations}
\begin{align}
&\tilde{q} \to 0: \qquad - \tilde{f}_{\theta} \le \tilde{g}_{\kappa}  \qquad \text{and} \qquad \tilde{f}_{\theta} \le 1 \;, \\
&\tilde{q} \to \infty: \qquad -1 \le \tilde{g}_{\kappa} \qquad \text{and} \qquad \tilde{f}_{\kappa} \le \tilde{g}_{\kappa} \;.
\end{align}
\end{subequations}
However, for $f_{\epsilon} < 0$ the inequalities that are found are different. Namely, they now read
\begin{subequations}
\begin{align}
&\tilde{q} \to 0: \qquad - \tilde{f}_{\theta} \le \tilde{g}_{\kappa}  \qquad \text{and} \qquad \tilde{f}_{\theta} \ge 1 \;, \\
&\tilde{q} \to \infty: \qquad -1 \ge \tilde{g}_{\kappa} \qquad \text{and} \qquad \tilde{f}_{\kappa} \le \tilde{g}_{\kappa} \;.
\end{align}
\end{subequations}
We therefore see that $\tilde{f}_{\theta}$ controls the decay rate of oscillations at large length scales ($\tilde{q}$ small), while $\tilde{f}_{\kappa}$ controls the decay rate at short length scales ($\tilde{q}$ large).

\subsection{Noisy equations: Simple scaling analysis}
States that are stable in the deterministic limit could potentially be destabilized by noise. The angular fluctuations were studied in the main text. Here we present a simple analysis to directly find the scaling of $P_\theta$ for small $q$. For this, we start from the Fourier transform of Eqs.~\eqref{eq:LocalDimEqs} and find
\begin{subequations}
\begin{align}
&i \Omega  \hat u = - f_{\epsilon} \hat u - f_{\theta} \hat \theta/(iq) - f_{\kappa} \hat \theta - i \hat \chi_\epsilon/q \;, \\
&i\Omega \hat \theta = - g_{\epsilon} \hat u - g_{\theta} \hat \theta/(iq) - g_{\kappa} \hat \theta - i \hat\chi_\kappa/q \;,
\end{align}
\end{subequations}
and for simplicity we will consider the case where $f_i, g_{\epsilon} = 0$ such that the equations for $\hat u$ and $\hat \theta$ decouple. In the absence of feedback, from the second equation $\hat \theta \sim \hat\chi_\kappa/q$ such that $\langle \hat \theta \hat \theta^\dagger \rangle \sim D_w/q^2$ and therefore $P_\theta \sim q^{-2}$. If only $g_\kappa \neq 0$, we still have $\hat \theta \sim \hat\chi_\kappa/q$. However, if $g_\theta \neq 0$, the bending terms are irrelevant for small $q$ such that $\hat{\theta} \sim \hat \chi_\kappa$ and therefore $\langle \hat \theta \hat \theta^\dagger \rangle \sim D_w$ and $P_\theta \sim q^0$.

\section{Nonlocal Feedback}
We now consider the case of nonlocal feedback. In particular, we assume that signaling molecules are responsible for the nonlocal and time-delayed feedback. Following Ref.~\cite{al-mosleh2023}, signals are assumed to propagate diffusively such that the corresponding Green's function satisfies
\begin{equation}
\partial_t G_D(s,t) - D \partial_s^2 G_D(s,t) = D \delta(s) e^{-\Gamma t}\;,
\end{equation}
with $D$ a diffusion constant and $\Gamma^{-1}$ a delay timescale. Assuming that elastic equilibration is much faster than the propagation speed, we consider the quasistatic limit and the Fourier transforms of the overall (local and nonlocal) Green's functions are found to be
\begin{subequations}
\begin{align}
\hat{F}_i(q,\Omega) &= - f_i - \frac{\varphi_i}{\Gamma + i \Omega} \frac{1}{q^2} \;, \\
\hat{G}_i(q,\Omega) &= - g_i - \frac{\gamma_i}{\Gamma + i \Omega} \frac{1}{q^2}\;.
\end{align}
\end{subequations}

\subsection{Deterministic case}
\begin{figure*}
\centering
\includegraphics[width=\textwidth]{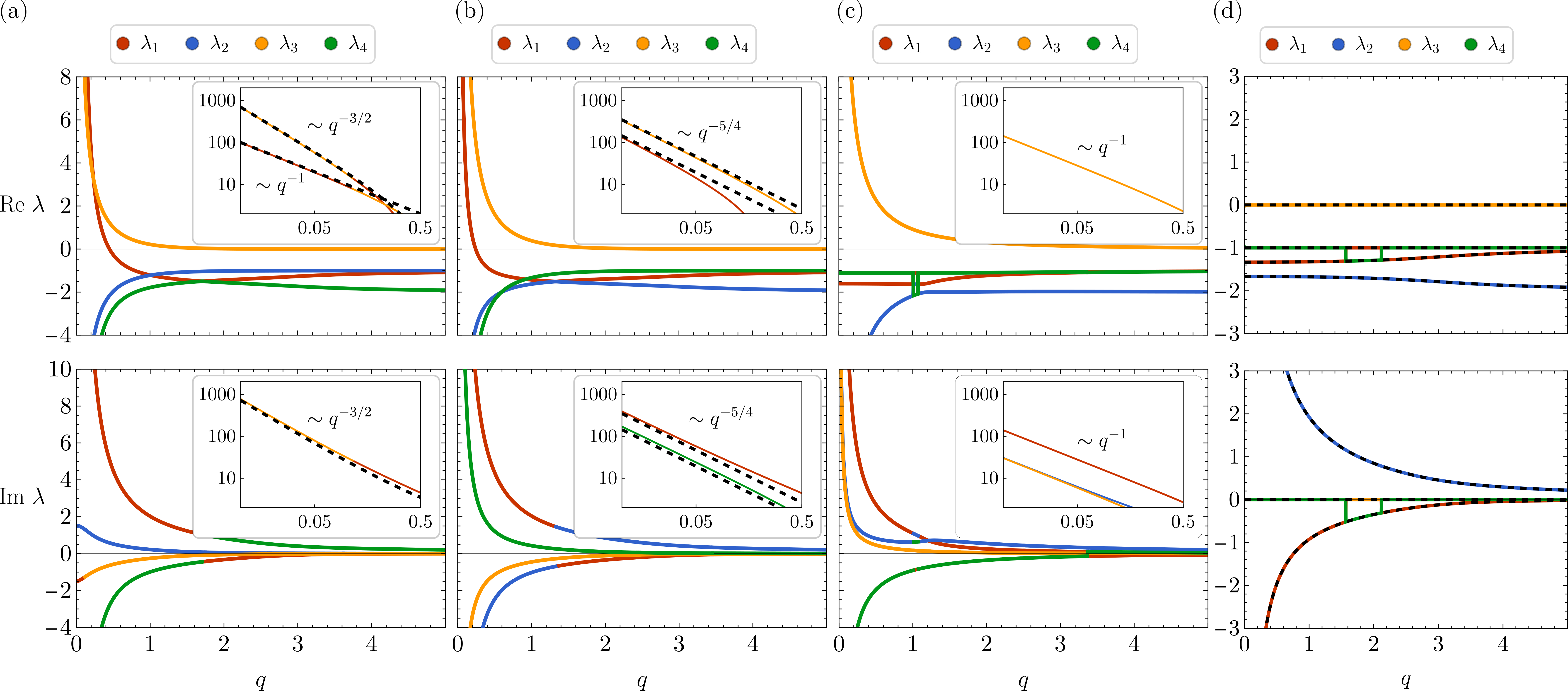}
\caption{\textbf{Diverging and finite solutions for nonlocal feedback.} (a) Real (top) and imaginary (bottom) parts of the eigenvalues found from solving Eq.~\eqref{eq:NonLocEigenval} with $f_i = g_i = \varphi_i = \gamma_i = 1$ for all $i$ and $\Gamma = 1$. The inset shows the respective curves diverging to $\infty$ as $q \to 0$, with the black dashed curves showing the asymptotic expressions in Eq.~\eqref{eq:NonLocGamma7}. (b) The same curves as in (a), but with $\gamma_{\theta} = 0$. The black dashed curves in the inset correspond to Eq.~\eqref{eq:NonLocGamma3}. (c) The same curves but with $\varphi_{\theta} = \gamma_{\theta} = 0$ and $\gamma_{\kappa} = -2$. The analytical expression for $q \to 0$ is too lengthy to write explicitly, but the divergence is now found to be $\sim q^{-1}$. (d) We find that for $\varphi_{\theta} = \gamma_{\theta} = 0$, $f_i = g_i = \varphi_i = \gamma_i = 1$ for all other $i$, and $\Gamma = 1$ the real part of the eigenvalue is finite as $q \to 0$. It is possible to find a simple explicit expression for the eigenvalues (black dashed lines), namely Eq.~\eqref{eq:EqualMagQ0}, since we are considering the simple case where $f_i = g_i = g$ and $\varphi_i = \gamma_i = \gamma$ for all parameters except $\varphi_{\theta}, \gamma_{\theta}$.}
\label{fig:NonlocalGen}
\end{figure*}
\subsubsection{Eigenvalue formulas}
As before, in the absence of noise we can write the equations compactly as $[\lambda \bm{1} - \bm{K}] \cdot \bm{V} = 0$, where now
\begin{equation}
\bm{K}(q, \lambda) = \begin{pmatrix} - f_{\epsilon} - \frac{\varphi_{\epsilon}}{(\Gamma + \lambda) q^2} & \frac{i f_{\theta}}{q} + \frac{i\varphi_{\theta}}{(\Gamma + \lambda) q^3} - f_{\kappa} - \frac{ \varphi_{\kappa}}{(\Gamma + \lambda) q^2}  \\ - g_{\epsilon} - \frac{\gamma_{\epsilon}}{(\Gamma + \lambda) q^2} & \frac{i g_{\theta}}{q} + \frac{i \gamma_{\theta}}{(\Gamma + \lambda) q^3} - g_{\kappa} - \frac{\gamma_{\kappa}}{(\Gamma + \lambda) q^2} \end{pmatrix} \;,
\end{equation}
and we can find the eigenvalues by solving
\begin{equation}
\det\left[\lambda \mathbf{1} - \bm{K}(q,\lambda)\right] = 0 \;.
\label{eq:NonLocEigenval}
\end{equation}
However, since $\bm{K}(q,\lambda)$ now depends explicitly on $\lambda$, the resulting expressions for the four eigenvalues are too complex to write down, although they can in principle be obtained analytically. See Fig.~\ref{fig:NonlocalGen} for some examples.

\subsubsection{Asymptotic limits of eigenvalues}
For $q \to \infty$, the expressions simplify significantly, and the solutions read:
\begin{subequations}
\begin{equation}
\lambda_{1,2,3,4} = \frac{1}{4}\left[-f_{\epsilon} - g_{\kappa} - 2\Gamma + \mathcal{C}_{1,2} \mp \sqrt{\left(f_{\epsilon} + g_{\kappa} - 2 \Gamma - \mathcal{C}_{1,2}\right)^2}\right] + \mathcal{O}\left(q^{-1}\right) \;,
\end{equation}
where
\begin{equation}
\mathcal{C}_{1,2} = \mp \sqrt{4 f_{\kappa} g_{\epsilon} + (g_{\kappa} - f_{\epsilon})^2} \;.
\end{equation}
\end{subequations}

We find that $\text{Re }\lambda_{1,2,3,4} \le 0$ if the same conditions as above are fulfilled, namely $-f_{\epsilon} \le g_{\kappa}$ and $f_{\kappa} g_{\epsilon} \le g_{\kappa} f_{\epsilon}$ and in addition $\Gamma \ge 0$ is required. If $\Gamma < 0$ the system is always unstable. Thus, $\Gamma$ can influence the value of $\text{Re }\lambda_{1,2,3,4}$ for $q \to \infty$ but, as long as $\Gamma \ge 0$, the stability is not affected by the nonlocal contributions at short length scales.
The more interesting case is the limit of large length scales. We find for $q \to 0$ that the eigenvalues are given by
\begin{subequations}
\begin{align}
\lambda_{1,2} &= \pm \frac{\sqrt{\gamma_{\theta}(\bar{\varphi}_\theta - \varphi_{\epsilon} \gamma_{\theta})}}{\gamma_{\theta} q} + \mathcal{O}\left(q^{0}\right) \;, \\
\lambda_{3,4} &= \pm \left[ \frac{\sqrt{i \gamma_{\theta}}}{q^{3/2}} - i \frac{\bar{\varphi}_\theta + \gamma_{\theta} \gamma_{\kappa}}{2 q^{1/2} (i \gamma_{\theta})^{3/2}}\right] + \mathcal{O}\left(q^{0}\right) \;,
\label{eq:NonLocGamma7}
\end{align}
\end{subequations}
where we have defined $\bar{\varphi}_i = \varphi_i \gamma_{\epsilon}$. From this we see that $\text{Re } \lambda_{3} \sim q^{-3/2}$ always, for all values $\gamma_{\theta} \neq 0$ (see Fig.~\ref{fig:NonlocalGen}(a)). Thus, in order to obtain a stable system, $\gamma_{\theta} = 0$ is required. In this case the above expression for the eigenvalues is no longer valid. Instead, we find
\begin{align}
\lambda_{1,2,3,4} &= \pm\sqrt{\pm\frac{\bar{\varphi}_\theta}{\sqrt{i \bar{\varphi}_\theta}}} \frac{1}{q^{5/4}} + \mathcal{O}\left(q^{-3/4}\right) \;.
\label{eq:NonLocGamma3}
\end{align}
This expression also diverges, but with a different exponent, namely $\text{Re}\,\lambda_{1,3} \sim q^{-5/4}$ (see Fig.~\ref{fig:NonlocalGen}(b)). Thus, $\varphi_{\theta} \to 0$ is required as well. In this case there are regions in parameter space where $\text{Re } \lambda < 0$; however, due to the complexity of the expressions, it is not feasible to write down the inequalities determining the region of stability as before (see Fig.~\ref{fig:NonlocalGen}(c,d)). Instead, we consider two simpler cases for which explicit expressions can be written down: \textit{(i)} nonlocal feedback dominates such that $f_i, g_i \to 0$ and \textit{(ii)} the magnitude of the feedback parameters is equal, i.e., $f_i = g_i \equiv g$ and $\varphi_i = \gamma_i \equiv \gamma$.

\begin{figure*}
\centering
\includegraphics[width=\textwidth]{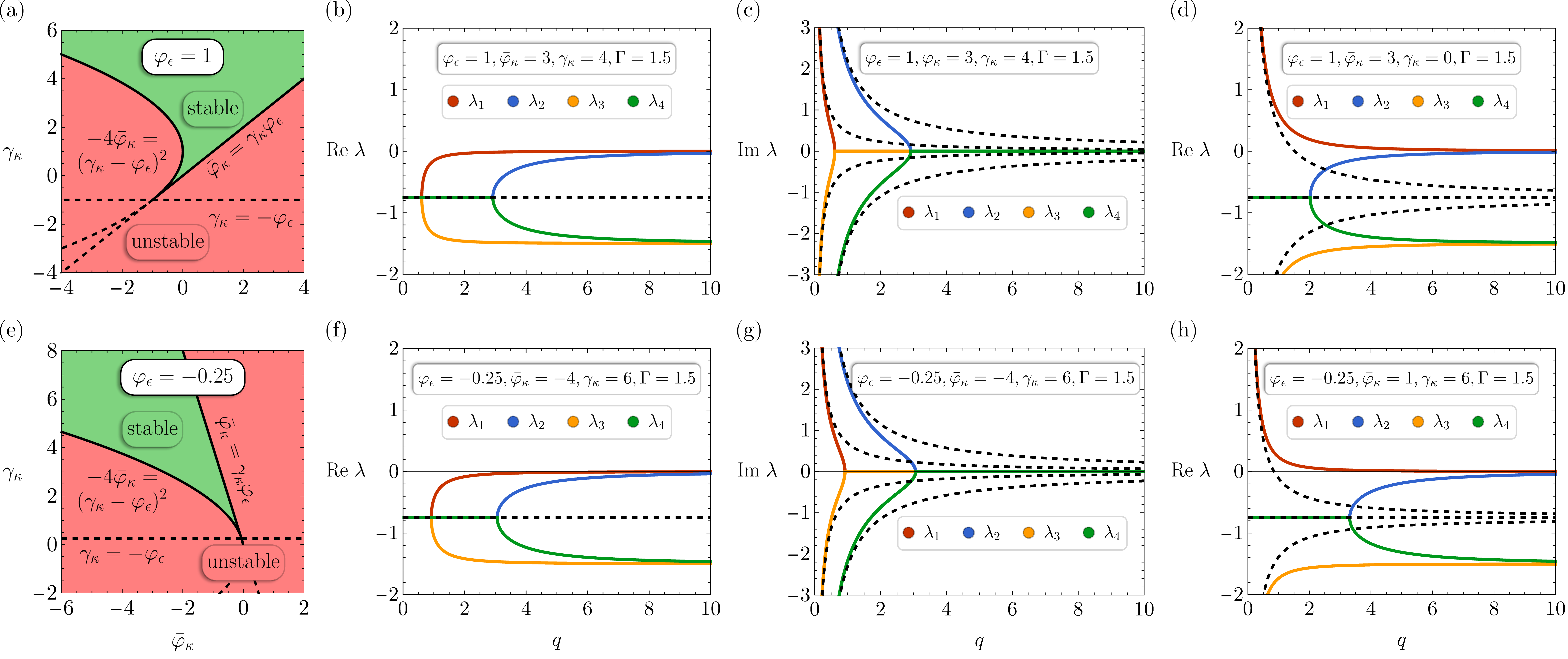}
\caption{\textbf{Stability for nonlocal feedback only.} (a) Stability diagram for the case of only nonlocal feedback ($f_i,\,g_i = 0$) for $\varphi_{\epsilon} = 1$. The region of stability is determined by the inequalities found from Eq.~\eqref{eq:OnlyNonLocalQ0}. (b) Example of the real and (c) imaginary parts of the four eigenvalues for the parameters $\varphi_{\epsilon} = 1$, $\bar{\varphi}_{\kappa} = 3$, $\gamma_{\kappa} = 4$, and $\Gamma = 1.5$. This parameter set lies in the stable region of (a) and thus $\text{Re } \lambda < 0$ for $q \to 0$ and furthermore $\Gamma > 0$ and thus $\text{Re } \lambda < 0$ for $q \to \infty$. The colored lines correspond to Eq.~\eqref{eq:OnlyNonLocalEV} while the black dashed lines correspond to Eq.~\eqref{eq:OnlyNonLocalQ0}. (d) Now, $\gamma_{\kappa} = 0$, i.e., the unstable region of parameter space, and thus $\text{Re } \lambda_1 \to \infty$ as $q \to 0$. (e) Stability diagram for $\varphi_{\epsilon} = -0.25$. (f) and (g) show the real and imaginary parts of the eigenvalues for a stable point ($\varphi_{\epsilon} = -0.25$, $\bar{\varphi}_{\kappa} = -4$, $\gamma_{\kappa} = 6$, and $\Gamma = 1.5$) while (h) shows an unstable point ($\bar{\varphi}_{\kappa} = 1$ now).}
\label{fig:OnlyNonlocal}
\end{figure*}

\textbf{Only nonlocal feedback.} If we set $f_i, g_i = 0$ and keep $\varphi_{\theta} = \gamma_{\theta} = 0$, but all other $\varphi_i, \gamma_i \neq 0$, we find that the eigenvalues are given by 
\begin{equation}
\lambda_{1,2,3,4} = -\frac{\Gamma}{2} \pm \sqrt{\frac{\Gamma^2}{4} -\frac{\gamma_{\kappa} + \varphi_{\epsilon} \pm i\sqrt{- 4 \bar{\varphi}_{\kappa} - (\gamma_{\kappa} - \varphi_{\epsilon})^2}}{2 q^2}} \;,
\label{eq:OnlyNonLocalEV}
\end{equation}
which for $q \to 0$ yields:
\begin{equation}
\lambda_{1,2,3,4} = -\frac{\Gamma}{2} \pm \sqrt{-\frac{\gamma_{\kappa} + \varphi_{\epsilon} \pm i\sqrt{- 4 \bar{\varphi}_{\kappa} - (\gamma_{\kappa} - \varphi_{\epsilon})^2}}{2 q^2}} + \mathcal{O}(q) \;.
\label{eq:OnlyNonLocalQ0}
\end{equation}
From this we find that the region inside the parabola bounded by the curves $\gamma_{\kappa} = \varphi_{\epsilon} \pm 2 i \sqrt{\bar{\varphi}_{\kappa}}$ is always unstable such that the region of stability is confined to the regions $\gamma_{\kappa} \le \varphi_{\epsilon} + 2 i \sqrt{\bar{\varphi}_{\kappa}}$ and $\gamma_{\kappa} \ge \varphi_{\epsilon} - 2 i \sqrt{\bar{\varphi}_{\kappa}}$ where these are real, i.e., for $\bar{\varphi}_{\kappa} \le 0$. Furthermore, similar to what we found for local feedback above, $\gamma_{\kappa} \varphi_{\epsilon} \ge \bar{\varphi}_{\kappa}$ defines a line that bounds the region of stability. Finally, $\gamma_{\kappa} \ge -\varphi_{\epsilon}$; geometrically, this condition constrains $\gamma_{\kappa}$ to values greater than the point of contact between the parabola and the line. See Fig.~\ref{fig:OnlyNonlocal} for the stability diagrams and some example curves of the eigenvalues.

\begin{figure*}
\centering
\includegraphics[width=\textwidth]{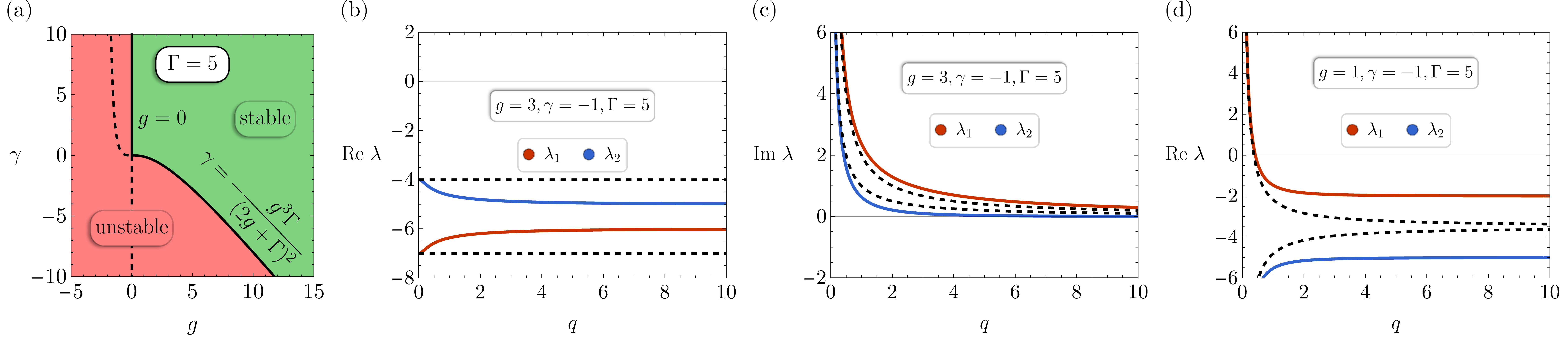}
\caption{\textbf{Stability for feedback with equal magnitude.} (a) Stability diagram for the case where both local and nonlocal feedback are present, but $f_i = g_i = g$ and $\varphi_i = \gamma_i = \gamma$ for all $i$ except $\varphi_{\theta} = \gamma_{\theta} = 0$. The stable region is found from Eq.~\eqref{eq:EqualMagQ0}. (b) Real and (c) imaginary part of the two nontrivial eigenvalues for a stable point ($g = 3$, $\gamma = -1$) with colored curves given by Eq.~\eqref{eq:EqualMagEV}, while the black dashed curves show the lowest-order expression for $q \to 0$, i.e., Eq.~\eqref{eq:EqualMagQ0}. (d) Unstable case with $g = 1$ now. $\Gamma = 5$ everywhere here.
}
\label{fig:EqualMag}
\end{figure*}

\textbf{Equal magnitude.} Now, we set $f_i = g_i = g$ for all $i$ as well as $\varphi_i = \gamma_i = \gamma$, except we keep $\varphi_{\theta} = \gamma_{\theta} = 0$. In this case the two nontrivial eigenvalues are 
\begin{equation}
\lambda_{1,2} = -\frac{\Gamma + 2g}{2} + \frac{i g \pm i \sqrt{8 \gamma - (i g - 2 g q + q \Gamma)^2}}{2q}
\label{eq:EqualMagEV}
\end{equation}
which in the limit $q \to 0$ becomes
\begin{equation}
\lambda_{1,2} = -\frac{\Gamma + 2g}{2} \pm g \frac{\Gamma - 2g}{2 \sqrt{g^2 + 8 \gamma}} + i \frac{g \pm \sqrt{g^2 + 8\gamma}}{2q} + \mathcal{O}(q) \;.
\label{eq:EqualMagQ0}
\end{equation}
Considering their real parts, we find from the last term that there is again a parabola in parameter space inside of which the system is always unstable; thus $\gamma \ge -g^2/8$ is required for stability such that the diverging term $\sim q^{-1}$ is always imaginary. Furthermore, requiring the terms $\sim q^0$ to be nonpositive gives the conditions $\gamma \ge g^3 \Gamma/(2 g + \Gamma)^2$ and $g \ge -\Gamma/2$. However, from the conditions of stability at $q \to \infty$ we also have $g, \Gamma \ge 0$, which are more restrictive than $g \ge -\Gamma/2$. This also implies that $g^3 \Gamma/(2 g + \Gamma)^2 \ge -g^2/8$ always such that the only two relevant conditions for stability are $g, \Gamma \ge 0$ and $\gamma \ge g^3 \Gamma/(2 g + \Gamma)^2$. See Fig.~\ref{fig:EqualMag} for the stability diagrams and some example curves of the eigenvalues.

\subsection{Noisy equations}
$P_\theta$ is found as in the local case (see main text). As the full expression for $P_{\theta}$ is very lengthy, we do not write it here. Instead, we consider the asymptotic expansion. As before, $P_{\theta} \to 0$ as $q \to \infty$. As described in the main text, when taking $q \to 0$ we find
\begin{equation}
P_{\theta} = \frac{D_w \varphi_{\epsilon}^2 + D_u \gamma_{\epsilon}^2}{\left(\varphi_{\theta} \gamma_{\epsilon} - \varphi_{\epsilon} \gamma_{\theta}\right)^2} \left( \Gamma^2 + \Omega^2\right) q^4 + \mathcal{O}\left(q^{6}\right) \;,
\end{equation}
whereas, after taking the limits $\gamma_{\theta} \to 0$ and $\varphi_{\theta} \to 0$ (to guarantee stability in the deterministic limit) we find
\begin{equation}
P_{\theta} = \frac{D_w \varphi_{\epsilon}^2 + D_u \gamma_{\epsilon}^2}{\left(\varphi_{\kappa} \gamma_{\epsilon} - \varphi_{\epsilon} \gamma_{\kappa}\right)^2} \left( \Gamma^2 + \Omega^2\right) q^2 + \mathcal{O}\left(q^{4}\right) \;.
\end{equation}
To systematically identify the minimal parameter requirements for obtaining a finite $P_\theta$ as $q \to 0$ we first consider the case where all but one of the parameters $\{f_i, g_i, \varphi_i,\gamma_i\}$ are taken to be zero. We find the leading-order terms shown in Tab.~\ref{tab:SinglePara}, where the three cases in which $P_\theta$ is finite as $q \to 0$, namely nonzero $g_{\theta}$, $\gamma_{\theta}$, or $\gamma_{\kappa}$, are highlighted in bold.
\begin{table}[h]
\centering
\renewcommand{\arraystretch}{1.2}
\begin{tabular}{| c | | c | c | c | c | c | c |}
\hline
$\displaystyle \text{Parameter}$ & $\displaystyle f_{\epsilon}$ & $\displaystyle f_{\theta}$ & $\displaystyle f_{\kappa}$ & $\displaystyle g_{\epsilon}$ & $\displaystyle \bm{g_{\theta}}$ & $\displaystyle g_{\kappa}$ \\
\hline
\rule[1ex]{0pt}{4ex}
$\displaystyle P_{\theta}\centerstrut{3.75ex}$ & $\displaystyle \displaystyle \frac{D_w}{\Omega^2 q^2}\centerstrut{3.75ex}$ & $\displaystyle \frac{D_w}{\Omega^2 q^2}$ & $\displaystyle \frac{D_w}{\Omega^2 q^2}\centerstrut{3.75ex}$ & $\displaystyle \frac{D_u g_{\epsilon}^2 + D_w \Omega^2}{q^2 \Omega^2}\centerstrut{3.75ex}$ & $\displaystyle \bm{\frac{D_w}{g_{\theta}^2}}\centerstrut{3.75ex}$ & $\displaystyle \frac{D_w}{\Omega^2 q^2}\centerstrut{3.75ex}$ \\
\hline
\hline
$\displaystyle \text{Parameter}$ & $\displaystyle \varphi_{\epsilon}$ & $\displaystyle \varphi_{\theta}$ & $\displaystyle \varphi_{\kappa}$ & $\displaystyle \gamma_{\epsilon}$ & $\displaystyle \bm{\gamma_{\theta}}$ & $\displaystyle \bm{\gamma_{\kappa}}$ \\
\hline
\rule[1ex]{0pt}{4ex}
$\displaystyle P_{\theta}\centerstrut{3.75ex}$ & $\displaystyle \frac{D_w}{\Omega^2 q^2}\centerstrut{3.75ex}$ & $ \displaystyle\frac{D_w}{\Omega^2 q^2}\centerstrut{3.75ex}$ & $\displaystyle \frac{D_w}{\Omega^2 q^2}$ & $\displaystyle \frac{D_u \gamma_{\epsilon}^2}{q^6(\Gamma^2 \Omega^4 + \Omega^6)}\centerstrut{3.75ex}$ & $\displaystyle \bm{\frac{D_w q^4 (\Gamma^2 + \Omega^2)}{\gamma_{\theta}^2}}\centerstrut{3.75ex}$ & $\displaystyle \bm{\frac{D_w q^2 (\Gamma^2 + \Omega^2)}{\gamma_{\kappa}^2}}\centerstrut{3.75ex}$ \\
\hline
\end{tabular}
\caption{Leading-order small-$q$ scaling of $P_\theta(q)$ for cases in which only one feedback parameter is nonzero, with finite cases highlighted in bold.}
\label{tab:SinglePara}
\end{table}
For the remaining parameters, we next consider cases in which the product of two parameters is nonzero. The results are shown in Tab.~\ref{tab:TwoParas}. All combinations of two parameters not shown in this table have the expression $D_w/(q^2(g_{\kappa}^2 + \Omega^2))$ (if $g_{\kappa} \neq 0$) or $D_w/(q^2 \Omega^2)$ (if $g_{\kappa} = 0$).

\begin{table}[h!]
\centering
\renewcommand{\arraystretch}{1.2}
\begin{tabular}{|c|c|c|c|c|}
\hline
$\displaystyle f_{\epsilon} g_{\epsilon}$ & $\displaystyle \bm{f_{\theta} g_{\epsilon}}$ & $\displaystyle f_{\kappa} g_{\epsilon}$ & $\displaystyle g_{\kappa} g_{\epsilon}$ & $\displaystyle \bm{\varphi_{\theta} g_{\epsilon}}$ \\
\hline
\rule[1ex]{0pt}{4ex}
$\displaystyle \frac{D_u g_{\epsilon}^2+D_w(f_{\epsilon}^2+\Omega^2)}{q^2 \Omega^2 (f_{\epsilon}^2+\Omega^2)}\centerstrut{3.75ex}$ &
$\displaystyle \bm{\frac{D_u g_{\epsilon}^2+D_w \Omega^2}{f_{\theta}^2 g_{\epsilon}^2}}\centerstrut{3.75ex}$ &
$\displaystyle \frac{D_u g_{\epsilon}^2+D_w \Omega^2}{q^2 (f_{\kappa} g_{\epsilon}+\Omega^2)^2}\centerstrut{3.75ex}$ &
$\displaystyle \frac{D_u g_{\epsilon}^2+D_w \Omega^2}{q^2 \Omega^2 (g_{\kappa}^2+\Omega^2)}\centerstrut{3.75ex}$ &
$\displaystyle \bm{\frac{q^4 (\Gamma^2+\Omega^2)(D_u g_{\epsilon}^2+D_w \Omega^2)}{\varphi_{\theta}^2 g_{\epsilon}^2}}\centerstrut{3.75ex}$ \\
\hline\hline
$\displaystyle \bm{\varphi_{\kappa} \gamma_{\epsilon}}$ & $\displaystyle f_{\epsilon} \gamma_{\epsilon}$ & $\displaystyle \bm{f_{\theta} \gamma_{\epsilon}}$ & $\displaystyle f_{\kappa} \gamma_{\epsilon}$ & $\displaystyle g_{\epsilon} \gamma_{\epsilon}$ \\
\hline
\rule[1ex]{0pt}{4ex}
$\displaystyle \bm{\frac{q^2 (\Gamma^2+\Omega^2)(D_u \gamma_{\epsilon}^2+D_w \Omega^2)}{\varphi_{\kappa}^2 \gamma_{\epsilon}^2}}\centerstrut{3.75ex}$ &
$\displaystyle \frac{D_u \gamma_{\epsilon}^2 }{q^6 \Omega^2 (\Gamma^2+\Omega^2) (f_{\epsilon}^2+\Omega^2)}\centerstrut{3.75ex}$ &
$\displaystyle \bm{\frac{D_u}{f_{\theta}^2}}\centerstrut{3.75ex}$ &
$\displaystyle \frac{D_u}{f_{\kappa}^2 q^2}\centerstrut{3.75ex}$ &
$\displaystyle \frac{D_u \gamma_{\epsilon}^2 }{q^6 \Omega^4 (\Gamma^2+\Omega^2)}\centerstrut{3.75ex}$ \\
\hline\hline
$\displaystyle g_{\kappa} \gamma_{\epsilon}$ & $\displaystyle \varphi_{\epsilon} \gamma_{\epsilon}$ & $\displaystyle \bm{\varphi_{\theta} \gamma_{\epsilon}}$ & $\displaystyle \bm{\varphi_{\kappa} \gamma_{\epsilon}}$ & \\
\hline
\rule[1ex]{0pt}{4ex}
$\displaystyle \frac{D_u \gamma_{\epsilon}^2 }{q^6 \Omega^2 (\Gamma^2+\Omega^2) (g_{\kappa}^2+\Omega^2)}\centerstrut{3.75ex}$ &
$\displaystyle \frac{D_u \gamma_{\epsilon}^2 + D_w \varphi_{\epsilon}^2 }{q^2 \varphi_{\epsilon}^2 \Omega^2}\centerstrut{3.75ex}$ &
$\displaystyle \bm{\frac{D_u q^4 (\Gamma^2+\Omega^2)}{\varphi_{\theta}^2}}\centerstrut{3.75ex}$ &
$\displaystyle \bm{\frac{D_u q^2 (\Gamma^2+\Omega^2)}{\varphi_{\kappa}^2}}\centerstrut{3.75ex}$ & \\
\hline
\end{tabular}
\caption{Leading-order small-$q$ scaling of $P_\theta(q)$ for minimal two-parameter combinations. Couplings which suppress the angular fluctuation divergence are highlighted in bold.}
\label{tab:TwoParas}
\end{table}

\section{Elastic substrate}
Combining Eqs.~\eqref{eq:EOMSI} and~\eqref{eq:GeneralFeedbackSI} for local feedback in the limit in which $f_{\theta} = f_{\kappa} = g_{\epsilon} = g_{\theta} = 0$ and $f_{\epsilon} = g_{\kappa} = g$ yields

\begin{subequations}
\begin{align}
&\partial_t\partial_s u + g \epsilon(s,t) - \chi_\epsilon(s,t) = 0 \;, \\
&B \partial^2_s \left(\partial_t \partial_s^2 w + g \kappa(s,t) - \chi_\kappa(s,t)\right) + k \partial_t w = 0 \;.
\end{align}
\end{subequations}
In the deterministic limit, substituting the ansatz $(u,w) = (u_0,w_0) e^{i q s + \lambda t}$ yields
\begin{subequations}
\begin{align}
&\left[\lambda + g\right] u = 0 \;, \\
&\left[\left(\lambda + g\right) q^4 + \lambda \tilde{k}\right] w = 0 \;,
\end{align}
\end{subequations}
which can be written in the form $\left[\lambda \bm{1} - \bm{K} \right]\cdot \bm{V} = 0$ with
\begin{equation}
\bm{K} = \begin{pmatrix} -g & 0 \\ 0 & \frac{-g q^4}{\tilde{k} + q^4}\end{pmatrix} \;.
\end{equation}
The eigenvalues are thus simply the diagonal entries of $\bm{K}$. In the presence of noise, the Fourier transform of the relevant equation (as the equations for $\hat u$ and $\hat w$ decouple, we only need to consider the latter) reads 
\begin{equation}
i \Omega q^4 \hat{w} + g q^4 \hat{w} + q^2 \hat{\chi}_\kappa(q,\Omega) + i \Omega \tilde{k} \hat{w} = 0 \Rightarrow \hat w(q,\Omega) = -\frac{q^2 \hat\chi_\kappa}{g q^4 + i (\tilde{k}+q^4)\Omega}
\end{equation}
and from this we find
\begin{equation}
\left\langle \hat \theta(q,\Omega) \hat \theta^\dagger(q',\Omega')\right\rangle \sim q q' \left\langle \hat w(q,\Omega) \hat w^\dagger(q',\Omega')\right\rangle \sim \frac{q^6 D_w}{g^2 q^8 + (\tilde{k} + q^4)^2\Omega^2} \;.
\end{equation}

\bibliography{Biblio.bib}